\documentclass[twoside,11pt]{article}

\usepackage{jmlr2e}

\usepackage{url}

\usepackage{mathtools}
\usepackage{graphicx}
\usepackage{amsmath}
\usepackage{amssymb}
\usepackage{enumitem}
\usepackage{algorithm}
\usepackage{multirow}
\usepackage{vhistory}

\usepackage{booktabs}

\usepackage{natbib}

\usepackage{epstopdf}

\usepackage{color}

\usepackage{mdframed}

\usepackage{listings}
\definecolor{dkgreen}{rgb}{0,0.6,0}
\definecolor{gray}{rgb}{0.5,0.5,0.5}
\definecolor{mauve}{rgb}{0.58,0,0.82}
\usepackage{siunitx}
\usepackage{booktabs}		
\usepackage{arydshln}		
\usepackage{algorithm}
\usepackage{algorithmic}
\usepackage{enumitem}
\usepackage{color}
\usepackage{multirow}
\usepackage{amsmath}
\usepackage{amssymb}
\usepackage{url}

\usepackage{natbib}

\usepackage{cleveref}  

\newcommand{\argmin}{\operatornamewithlimits{arg\,min}}

\newcommand{\NAcell}{\multicolumn{2}{c}{\scriptsize$\mbox{N/A}$}}

\hypersetup{pdfinfo={
Title={Bibliographic Analysis on Research Publications using Authors, Categorical Labels and the Citation Network},
Author={Kar Wai Lim, Wray Buntine},
Subject={Machine Learning},
Keywords={Bibliographic analysis, Topic model, Bayesian nonparametric, Author-citation network, Pitman-Yor process}
}}

\jmlrheading{103}{2016}{185--213}{24 Mar 2015}{23 Feb 2016}{Lim and Buntine}

\ShortHeadings{Bibliographic Analysis using Authors, Categorical Labels and Citation Network}{Lim and Buntine}
\firstpageno{1}

\begin{document}

\title{    
    Bibliographic Analysis on Research Publications 
    using Authors, Categorical Labels and the Citation Network
    }

\author{\name Kar Wai Lim \email karwai.lim@anu.edu.au \\
       \addr The Australian National University and NICTA, Australia
       \AND
       \name Wray Buntine \email wray.buntine@monash.edu \\
       \addr Monash University, Australia
        }

\editor{Hang Li, Dinh Phung, Tru Cao, Tu-Bao Ho, and Zhi-Hua Zhou}

\maketitle

\begin{abstract}
Bibliographic analysis considers the author's research areas, the 
citation network and the paper content among other things. In this 
paper, we combine these three in a topic model that produces a 
bibliographic model of authors, topics and documents, using a 
nonparametric extension of a combination of the Poisson mixed-topic 
link model and the author-topic model. 
This gives rise to the Citation Network Topic Model (CNTM).
We propose a novel and 
efficient inference algorithm for the CNTM to explore subsets of 
research publications from CiteSeer$^\mathrm{X}$. 
The publication datasets are 
organised into three corpora, totalling to about 168k publications with 
about 62k authors. The queried datasets are made available online. In 
three publicly available corpora in addition to the queried datasets, 
our proposed model demonstrates an improved performance in both model 
fitting and document clustering, compared to several baselines.
Moreover, our model allows extraction of additional useful knowledge 
from the corpora, such as the visualisation of the author-topics 
network.
Additionally, we propose a simple method to incorporate 
supervision into topic modelling to achieve further improvement
on the clustering task.%
\end{abstract}

\begin{keywords}%
Bibliographic analysis, Topic model, Bayesian nonparametric, 
Author-citation network, Pitman-Yor process
\end{keywords}

\section{Introduction}

Models of bibliographic data need to consider many kinds of information. 
Articles are usually accompanied by metadata such as authors, 
publication data, categories and time. Cited papers can also be 
available. When authors' topic preferences are modelled, we need to 
associate the document topic information somehow with the authors'. 
Jointly modelling text data with citation network information can be 
challenging for topic models, and the problem is confounded when also 
modelling author-topic relationships.

In this paper, we propose a topic model to jointly model authors' topic 
preferences, text content%
\footnote{Abstract and publication title.} 
and the citation network. The model is a nonparametric extension of 
previous models discussed in \Cref{sec:relatedwork}. 
Using simple assumptions and approximations, we derive a 
novel algorithm that allows the probability vectors in the model to be 
integrated out.  This yields
a Markov chain Monte Carlo (MCMC) inference \textit{via} discrete sampling. 

As an extension of our previous work \citep{LimBuntineCNTM}, we propose 
a supervised approach to improve document clustering, by making use of 
categorical information that is available. Our method allows the level 
of supervision to be adjusted through a variable, giving us a model with 
no supervision, semi-supervised or fully supervised. Additionally, we 
present a more extensive qualitative analysis of the learned topic 
models, and display a visualisation snapshot of the learned 
author-topics network. We also perform additional diagnostic tests to 
assess our proposed topic model.  For example, we study the convergence 
of the proposed learning algorithm and report on the 
computation complexity of the~algorithm.

In the next section, we discuss the related work. 
\Cref{sec:model},~\ref{sec:likelihood} and~\ref{sec:inference} 
detail our topic model and its inference algorithm. 
We describe the datasets in \Cref{sec:data} 
and report on experiments in \Cref{sec:experiment}. 
Applying our model on 
research publication data, we demonstrate the model's improved 
performance, on both model fitting and a clustering task, compared to 
several baselines. Additionally, 
in \Cref{sec:qualitative_analysis}, we qualitatively analyse the 
inference results produced by our model. We find that the learned 
topics have high comprehensibility. 
Additionally, we present a visualisation snapshot of the 
learned topic models.
Finally, we perform diagnostic 
assessment of the topic model in \Cref{sec:diagnostic} and 
conclude the paper in \Cref{sec:conclusion}.

\section{Related Work}
\label{sec:relatedwork}

Latent Dirichlet Allocation (LDA) \citep{Blei:2003:LDA:944919.944937} is 
the simplest Bayesian topic model used in modelling text, which also 
allows easy learning of the model. \cite{TehJor2010a} proposed the {\em 
Hierarchical Dirichlet process} (HDP) LDA, which utilises the Dirichlet 
process (DP) as a nonparametric prior which allows a non-symmetric, 
arbitrary dimensional topic prior to be used. Furthermore, one can 
replace the Dirichlet prior on the word vectors with the 
\textit{Pitman-Yor Process} (PYP, also known as the two-parameter 
Poisson Dirichlet process) \citep{Teh:2006:HBL:1220175.1220299}, which 
models the power-law of word frequency distributions in natural language 
\citep{Goldwater:2011:PPD:1953048.2021075}, yielding significant 
improvement \citep{Sato:2010:TMP:1835804.1835890}.

Variants of LDA allow incorporating more aspects of a particular task 
and here we consider authorship and citation information. The 
\textit{author-topic model} (ATM) 
\citep{Rosen-Zvi:2004:AMA:1036843.1036902} uses the authorship 
information to restrict topic options based on author. Some recent work 
jointly models the document citation network and text content. This 
includes the  \textit{relational topic model} 
\citep{chang2010hierarchical}, the \textit{Poisson mixed-topic link 
model} (PMTLM) \citep{ZhuYGM:2013} and \textit{Link-PLSA-LDA} 
\citep{Nallapati:2008:JLT:1401890.1401957}. An extensive review of these 
models can be found in \cite{ZhuYGM:2013}. The \textit{Citation Author 
Topic} (CAT) model \citep{Tu:2010:CAT:1944566.1944711} models the 
author-author network on publications based on citations using an 
extension of the ATM. Note that our work is different to CAT in that we 
model the author-document-citation network instead of author-author 
network.

The \textit{Topic-Link LDA} \citep{Liu:2009:TLJ:1553374.1553460} 
jointly models author and text by using the distance between the 
document and author topic vectors.
Similarly the Twitter-Network topic model \citep{Lim2013Twitter} 
models the author network%
\footnote{The author network here corresponds to the Twitter follower network.}
based on author topic 
distributions, but using a Gaussian process to model the network.
Note that our work considers the author-document-citation of 
\cite{Liu:2009:TLJ:1553374.1553460}. 
We use the PMTLM of \cite{ZhuYGM:2013} 
to model the network, which lets one integrate PYP
hierarchies with the PMTLM using efficient MCMC sampling.

There is also existing work on analysing the degree of authors' 
influence. On publication data, \cite{Kataria:2011:CST:2283696.2283777} 
and \cite{Mimno:2007:MDL:1255175.1255196} analyse influential authors 
with topic models, while \cite{Weng:2010:TFT:1718487.1718520}, 
\cite{Tang:2009:SIA:1557019.1557108}, and 
\cite{Liu:2010:MTI:1871437.1871467} use topic models to analyse users' 
influence on social media.

\section{Supervised Citation Network Topic Model}
\label{sec:model}

In our previous work \citep{LimBuntineCNTM}, we proposed the Citation 
Network Topic Model (CNTM) that jointly models the \textit{text}, 
\textit{authors}, and the \textit{citation network} of research 
publications (documents). The CNTM allows us to both model the authors 
and text better by exploiting the correlation between the authors and 
their research topics. However, the benefit of the above modelling is 
not realised when the author information is simply missing from the 
data. This could be due to error in data collection (\textit{e.g.}\ 
metadata not properly formatted), or even simply that the author 
information is lost during preprocessing.

In this section, we propose an extension of the CNTM that remedies the 
above issue, by making use of additional metadata that is available. For 
example, the metadata could be the research areas or keywords associated 
with the publications, which are usually provided by the authors during 
the publication submission. However, this information might not always 
be reliable as it is not standardised across different publishers or 
conferences. In this paper, rather than using the mentioned metadata, we 
will instead incorporate the categorical labels that were previously 
used as ground truth for evaluation. As such, our extension gives rise 
to a supervised model, which we will call the Supervised Citation 
Network Topic Model~(SCNTM).

\begin{figure}[t]
	\begin{center}
	\centerline{\includegraphics[width=0.85\columnwidth]{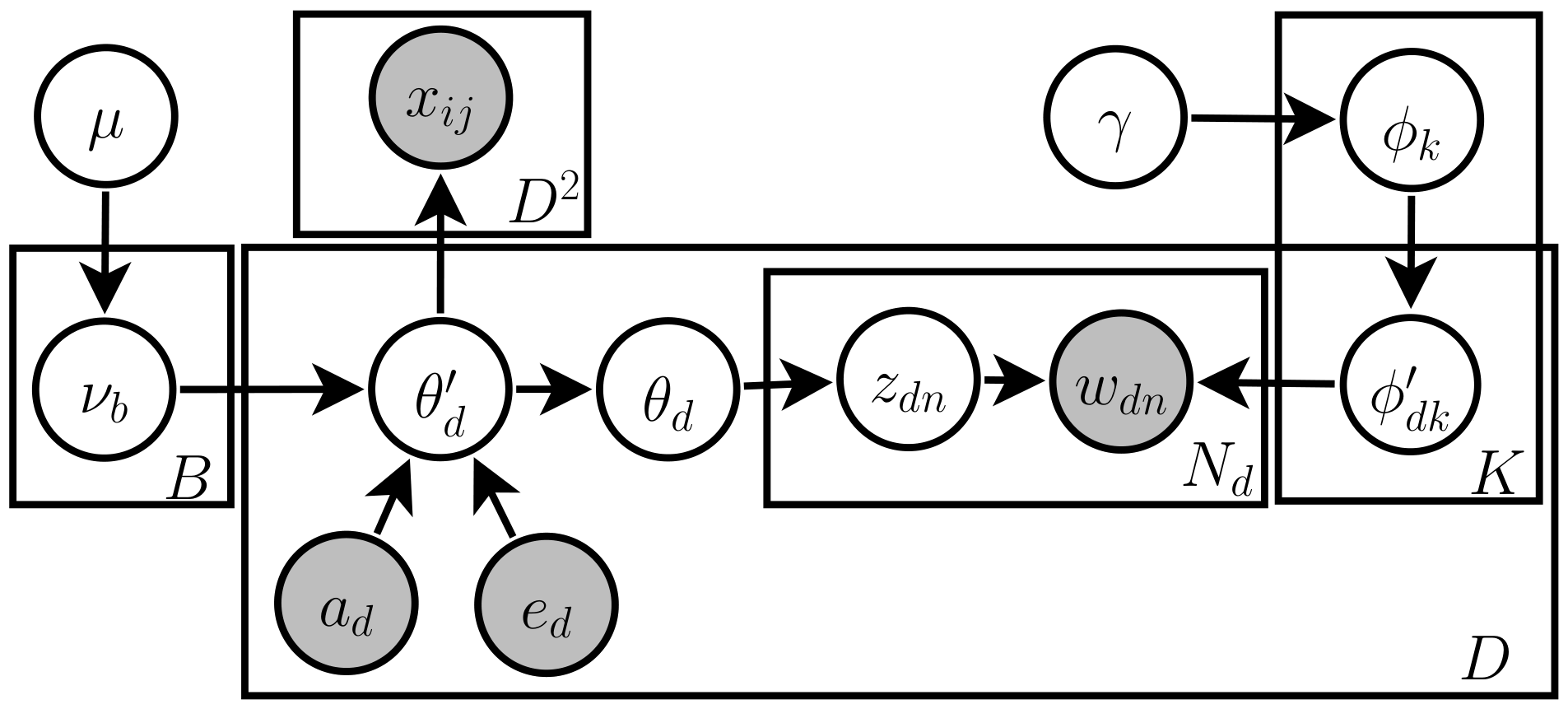} }
	\caption{Graphical model for SCNTM. The box on the top left with 
		$D^2$ entries is the citation network on documents 
		represented as a Boolean matrix. The remainder is a 
		nonparametric hierarchical PYP topic model where the
		labelled categories and authors are captured by the topic
		vectors $\nu$. The topic vectors $\nu$ influence the $D$ 
		documents' topic vectors $\theta'$ and $\theta$ based on the 
		observed authors $a$ or categories $e$. The latent topics and 
		associated words are represented by the variables $z$ and $w$.
		The $K$ topics, shown in the top right, have bursty modelling 
		following \cite{Buntine:2014:ENT:2623330.2623691}.
	}
	\label{fig:poissonNetworkModel}
	\end{center}
\end{figure}

We first describe the topic model part of SCNTM for which the citations 
are not considered, it will be used for comparison later in 
\Cref{sec:experiment}. We then complete the SCNTM with the 
discussion on its network component. The full graphical model for SCNTM 
is displayed in \Cref{fig:poissonNetworkModel}.

To clarify the notations used in this paper, \textit{variables that 
are without subscript represent a collection of variables of the 
same notation}. For instance, $w_d$ represents all the words in 
document $d$, that is, $w_d = \{w_{d1}, \dots, w_{dN_d}\}$ where 
$N_d$ is the number of words in document $d$; and $w$ represents all 
words in a corpus, $w=\{w_1, \dots, w_D\}$, where $D$ is the number 
of~documents.

\subsection{Hierarchical Pitman-Yor Topic Model}
\label{subsec:topic_model}

The SCNTM uses both the \textit{Griffiths-Engen-McCloskey} (GEM) 
distribution \citep{pitman1996some} and the {\it Pitman-Yor process} 
(PYP)~\citep{Teh:2006:HBL:1220175.1220299} to generate probability 
vectors. Both the GEM distribution and the PYP are parameterised by a 
\textit{discount} parameter $\alpha$ and a \textit{concentration} 
parameter $\beta$. The PYP is additionally parameterised by a 
\textit{base distribution} $H$, which is also the mean of the PYP when 
it can be represented by a probability vector. 
Note that the base distribution can also be a PYP. This gives rise 
to the hierarchical Pitman-Yor process (HPYP).

In modelling authorship, the SCNTM modifies the approach of the 
author-topic model \citep{Rosen-Zvi:2004:AMA:1036843.1036902} which 
assumes that the words in a publication are equally attributed to the 
different authors. This is not reflected in practice since publications 
are often written more by the first author, excepting when the order is 
alphabetical. Thus, we assume that the first author is dominant and 
attribute all the words in a publication to the first author. 
Although, we could model the contribution of each author on a 
publication by, say, using a Dirichlet distribution, we found that 
considering only the first author gives a simpler learning algorithm 
and cleaner results. 

The generative process of the topic model component of the SCNTM is 
as follows. We first sample a root topic distribution $\mu$ with a GEM 
distribution to act as a base distribution for the author-topic 
distributions $\nu_a$ for each author $a$, and also for the 
category-topic distributions $\nu_e$ for each category $e$:
\begin{align}
\mu &\sim \mathrm{GEM}(\alpha^\mu, \beta^\mu) ~, 
\\
\nu_a \,|\, \mu &\sim \mathrm{PYP}(\alpha^{\nu_a}, \beta^{\nu_a}, \mu) ~, 
&& a \in \mathcal{A} ~.
\\
\nu_e \,|\, \mu &\sim \mathrm{PYP}(\alpha^{\nu_e}, \beta^{\nu_e}, \mu) ~,
&& e \in \mathcal{E} ~.
\end{align}
Here, $\mathcal{A}$ represents the set of all authors while $\mathcal{E}$
denotes the set of all categorical labels in the text corpus. Note we 
have used the same symbol ($\nu$) for both the author-topic
distributions and the category-topic distributions.

We introduce a parameter $\eta$ called the \textit{author threshold} 
which controls the level of supervision used by SCNTM. We say an author 
$a$ is significant if the author has produced more than or equal to 
$\eta$ publications, \textit{i.e.}
\begin{align}
\mathrm{significance}(a) =
\left\{
	\begin{array}{ll}
		1  & ~~~\mathrm{if\ } \sum_d I(a_d = a) \geq \eta \\
		0  & ~~~\mathrm{otherwise. } 
	\end{array}
\right.
\end{align}
Here,
$a_d$ represents the author for document $d$, and $I(\triangle)$ is the 
indicator function that evaluates to $1$ if $\triangle$ is true, else 
$0$.

Next, for each document $d$ in a publication collection of size $D$, 
we sample the document-topic prior $\theta'_d$ from $\nu_{a_d}$ or 
$\nu_{e_d}$ depending on whether the author $a_d$ for the document is 
significant:
\begin{align}
\theta'_d \,|\, a_d,e_d,\nu \sim 
\left\{
	\begin{array}{ll}
		\mathrm{PYP}(\alpha^{\theta'_d}, \beta^{\theta'_d}, \nu_{a_d})   
		& ~~~\mathrm{if\ } \mathrm{significance}(a_d) = 1 
		\\
		\mathrm{PYP}(\alpha^{\theta'_d}, \beta^{\theta'_d}, \nu_{e_d})   
		& ~~~\mathrm{otherwise, } 
	\end{array}
\right.
&& ~~~~ d = 1, \dots, D ~,
\end{align}
where $e_d$ is the categorical label associated with document $d$.
For the sake of notational simplicity, we introduce a variable 
$b$ to capture both the author and the category. 
We let $b$ takes the value of $1, \dots, A$ for each author in $\mathcal{A}$,
and let $b$ takes the value of $(A+1), \dots, B$ for the 
categories in $\mathcal{E}$. Note that $B = |\mathcal{A}| + |\mathcal{E}|$.
Thus, we can also write the distribution of $\theta'_d$ as
\begin{align}
\theta'_d \,|\, \nu_b \sim \mathrm{PYP}(\alpha^{\theta'_d}, \beta^{\theta'_d}, \nu_b) 
&& ~~~~ d = 1, \dots, D ~,
\end{align}
where $b = a_d$ if $\mathrm{significance}(a_d) = 1$, else 
$b = e_d$\,.

By modelling this way, we are able to handle missing authors
and incorporate supervision into the SCNTM. For example, choosing 
$\eta = 1$ allows us to make use of the categorical information for 
documents that have no valid author. Alternatively, we could select a 
higher $\eta$, this smooths out the document-topic distributions for 
documents that are written by authors who have authored only a small 
number of publications. This treatment leads to a better clustering 
result as these authors are usually not discriminative enough for 
prediction. On the extreme, we can set $\eta = \infty$ to achieve full 
supervision. We note that the SCNTM reverts to the CNTM when $\eta = 0$, 
in this case the model is not supervised.


We then sample the document-topic distribution 
$\theta_d$ given $\theta'_d$:
\begin{align}
\theta_d \,|\, \theta'_d \sim 
\mathrm{PYP}(\alpha^{\theta_d}, \beta^{\theta_d}, \theta'_d) ~,
&& ~~~~~~ d = 1, \dots, D ~.
\end{align}
Note that instead of modelling a single document-topic distribution, 
we model a document-topic hierarchy with $\theta'$ and $\theta$.
The primed $\theta'$ represents the topics of the document in the 
context of the citation network. The unprimed $\theta$ represents 
the topics of the text, naturally related to $\theta'$ but not the 
same. Such modelling gives citation information a higher impact to 
take into account the relatively low amount of citations compared to the 
text. The technical details on the effect of such modelling is presented 
in \Cref{subsec:modelling_topic_hierarchy}.

For the vocabulary side, we generate a background word distribution 
$\gamma$ given $H^\gamma$, a discrete uniform vector of length 
$|\mathcal{V}|$, \textit{i.e.}\ 
$H^\gamma = (\cdots, \frac{1}{|\mathcal{V}|}, \cdots)$. 
$\mathcal{V}$ is the set of \textit{distinct} word tokens observed in a 
corpus. Then, we sample a topic-word distribution $\phi_k$ for each 
topic $k$, with $\gamma$ as the base distribution:
\begin{align}
\gamma \sim \mathrm{PYP}(\alpha^\gamma, \beta^\gamma, H^\gamma) ~, 
\\
\phi_k \,|\, \gamma \sim 
\mathrm{PYP}(\alpha^{\phi_k}, \beta^{\phi_k}, \gamma) ~,
&&
k = 1, \dots, K ~.
\end{align}
Modelling word burstiness \citep{Buntine:2014:ENT:2623330.2623691} 
is important since words in a document are likely to repeat in the 
document. The same applies to publication abstract, as shown in 
\Cref{sec:data}. To address this property, we make the topics 
bursty so each document only focuses on a subset of words in the topic. 
This is achieved by defining the document-specific topic-word 
distribution $\phi'_{dk}$ for each topic $k$ in document $d$ as:
\begin{align}
\phi'_{dk} \,|\, \phi_k \sim 
\mathrm{PYP}(\alpha^{\phi'_{dk}}, \beta^{\phi'_{dk}}, \phi_k) ~,
&&
d = 1, \dots, D ~,
~~~
k = 1, \dots, K ~.
\end{align}

\noindent
Finally, for each word $w_{dn}$ in document $d$, we sample the 
corresponding topic assignment $z_{dn}$ from the document-topic 
distribution $\theta_d$; while the word $w_{dn}$ is sampled from the 
topic-word distribution $\phi'_d$ given $z_{dn}$:
\begin{align}
z_{dn} \,|\, \theta_d \sim \mathrm{Discrete}(\theta_d) ~,
\\
w_{dn} \,|\, z_{dn}, \phi'_d \sim \mathrm{Discrete}(\phi'_{dz_{dn}}) ~,
&&
d = 1, \dots, D ~,
~~~
n = 1, \dots, N_d ~.
\end{align}
Note that $w$ includes words from the publications' title and abstract, 
but not the full article. This is because title and abstract provide a 
good summary of a publication's topics and thus more suited for topic 
modelling, while the full article contains too much technical detail 
that might not be too relevant.

In the next section, we describe the modelling of the citation network
accompanying a publication collections.  This completes the SCNTM. 

\subsection{Citation Network Poisson Model}
\label{subsec:network}

To model the citation network between publications, we assume that the 
citations are generated conditioned on the topic distributions $\theta'$ 
of the publications. Our approach is motivated by the degree-corrected 
variant of PMTLM \citep{ZhuYGM:2013}. Denoting $x_{ij}$ as the number of 
times document $i$ citing document $j$, we model $x_{ij}$ with a Poisson 
distribution with mean parameter $\lambda_{ij}$:
\begin{align}
x_{ij} \,|\, \lambda_{ij} & \sim  \mathrm{Poisson}(\lambda_{ij})~,
\nonumber \\
\lambda_{ij} & = 
\textstyle
\lambda_i^+ \lambda_j^- \sum_k \lambda^T_k \theta'_{ik} 
\theta'_{jk} ~,
&& 
i = 1, \dots, D ~, 
~~~
j = 1, \dots, D ~.
\label{eq:lambda_ij}
\end{align}
Here, $\lambda_i^+$ is the propensity of document $i$ to cite and 
$\lambda_j^-$ represents the popularity of cited document $j$, while 
$\lambda^T_k$ scales the $k$-th topic, effectively penalising common 
topics and strengthen rare topics. Hence, a citation from document $i$ 
to document $j$ is more likely when these documents are having relevant 
topics. Due to the limitation of the data, the $x_{ij}$ can only be $0$ 
or $1$, \textit{i.e.}\ it is a Boolean variable. Nevertheless, the 
Poisson distribution is used instead of a Bernoulli distribution because 
it leads to dramatically reduced complexity in analysis 
\citep{ZhuYGM:2013}. Note that the Poisson distribution is similar to 
the Bernoulli distribution when the mean parameter is small.
We present a list of variables
associated with the SCNTM in \Cref{tbl:variables}.

\begin{table}[t!]
    \centering
    \caption[
    	List of Variables for the Supervised Citation Network Topic Model
    ]{
    	List of Variables for the Supervised Citation Network Topic Model (SCNTM).
    }
    \label{tbl:variables}
    \vspace{2mm}
    \begin{tabular}{cp{0.21\columnwidth}p{0.595\columnwidth}}
    \toprule
    \multicolumn{1}{c}{Variable} 
    & \multicolumn{1}{c}{Name}
    & \multicolumn{1}{c}{Description} 
    \\
    \midrule
    \multirow{1}{*}{$z_{dn}$}
    & \multirow{1}{\linewidth}{\centering Topic}
    & Topical label for word $w_{dn}$\,.
    \\
    \noalign{\vskip 2.5pt} 
    \hdashline
    \noalign{\vskip 2.5pt}  
    \multirow{1}{*}{$w_{dn}$}
    & \multirow{1}{\linewidth}{\centering Word}
    & Observed word or phrase at position $n$ in document $d$.
    \\
    \noalign{\vskip 2.5pt} 
    \hdashline
    \noalign{\vskip 2.5pt}  
    \multirow{1}{*}{$x_{ij}$}
    & \multirow{1}{\linewidth}{\centering Citations}
    & Number of times document $i$ cites document $j$. 
    \\
    \noalign{\vskip 2.5pt} 
    \hdashline
    \noalign{\vskip 2.5pt}  
    \multirow{1}{*}{$a_{d}$}
    & \multirow{1}{\linewidth}{\centering Author}
    & Author for document $d$. 
    \\
    \noalign{\vskip 2.5pt} 
    \hdashline
    \noalign{\vskip 2.5pt}  
    \multirow{1}{*}{$e_d$}
    & \multirow{1}{\linewidth}{\centering Category}
    & Category label for document $d$. 
    \\
    \noalign{\vskip 2.5pt} 
    \hdashline
    \noalign{\vskip 2.5pt}  
    \multirow{2}{*}{$\phi'_{dk}$}
    & \centering Document-topic- word distribution
    & \multirow{2}{\linewidth}{%
        Probability distribution in generating words given document $d$ and topic $k$.
      }
    \\
    \noalign{\vskip 2.5pt} 
    \hdashline
    \noalign{\vskip 2.5pt}  
    \multirow{2}{*}{$\phi_k$}
    & \centering Topic-word distribution
    & \multirow{2}{\linewidth}{Word prior for $\phi'_{dk}$\,.}
    \\
    \noalign{\vskip 2.5pt} 
    \hdashline
    \noalign{\vskip 2.5pt}  
    \multirow{2}{*}{$\theta_d$}
    & \centering Document-topic distribution
    & \multirow{2}{\linewidth}{Probability distribution in generating topics for document $d$.}
    \\
    \noalign{\vskip 2.5pt} 
    \hdashline
    \noalign{\vskip 2.5pt}  
    \multirow{2}{*}{$\theta'_d$}
    & \centering Document-topic prior
    & \multirow{2}{\linewidth}{Topic prior for $\theta_d$\,.}
    \\
    \noalign{\vskip 2.5pt} 
    \hdashline
    \noalign{\vskip 2.5pt}  
    \multirow{2}{*}{$\nu_b$}
    & \centering Author/category- topic distribution
    & \multirow{2}{\linewidth}{
        Probability distribution in generating topics for author or category $b$.}
    \\
    \noalign{\vskip 2.5pt} 
    \hdashline
    \noalign{\vskip 2.5pt}  
    \multirow{2}{*}{$\gamma$}
    & \centering Global word distribution
    & \multirow{2}{\linewidth}{Word prior for $\phi_k$\,.}
    \\
    \noalign{\vskip 2.5pt} 
    \hdashline
    \noalign{\vskip 2.5pt}  
    \multirow{2}{*}{$\mu$}
    & \centering Global topic distribution
    & \multirow{2}{\linewidth}{Topic prior for $\nu_b$\,.}
    \\
    \noalign{\vskip 2.5pt} 
    \hdashline
    \noalign{\vskip 2.5pt}  
    $\alpha^\mathcal{N}$  
    & \centering Discount 
    & Discount parameter of the PYP $\mathcal{N}$. 
    \\
    \noalign{\vskip 2.5pt} 
    \hdashline
    \noalign{\vskip 2.5pt} 
    $\beta^\mathcal{N}$  
    & \centering Concentration 
    & Concentration parameter of the PYP $\mathcal{N}$. 
    \\
    \noalign{\vskip 2.5pt} 
    \hdashline
    \noalign{\vskip 2.5pt}     
    $H^\mathcal{N}$ 
    & \centering Base distribution 
    & Base distribution of the PYP $\mathcal{N}$. 
    \\
    \noalign{\vskip 2.5pt} 
    \hdashline
    \noalign{\vskip 2.5pt}  
    $\lambda_{ij}$  
    & \centering Rate 
    & Rate parameter or the mean for $x_{ij}$\,.
    \\
    \noalign{\vskip 2.5pt} 
    \hdashline
    \noalign{\vskip 2.5pt} 
    $\lambda_i^+$  
    & \centering Cite propensity
    & Propensity to cite for document $i$. 
    \\
    \noalign{\vskip 2.5pt} 
    \hdashline
    \noalign{\vskip 2.5pt}     
    $\lambda_i^-$ 
    & \centering Cited propensity
    & Propensity to be cited for document $j$.
    \\
    \noalign{\vskip 2.5pt} 
    \hdashline
    \noalign{\vskip 2.5pt}     
    $\lambda_k^T$ 
    & \centering Scaling factor
    & Citation scaling factor for topic $k$.
    \\   
    \bottomrule
    \end{tabular}
\end{table}

\section{Model Representation and Posterior Likelihood}
\label{sec:likelihood}

Before presenting the posterior used to develop the MCMC sampler, we 
briefly review handling of the hierarchical PYP models in 
\Cref{subsec:mhpyp}. We cannot provide an adequately detailed 
review in this paper, thus we present the main ideas.

\subsection{Modelling with Hierarchical PYPs}
\label{subsec:mhpyp}
The key to efficient sampling with PYPs is to marginalise out the 
probability vectors (\textit{e.g.}\ topic distributions) in the model 
and record various associated counts instead, thus yielding a collapsed 
sampler. While a common approach here is to use the hierarchical Chinese 
Restaurant Process (CRP) of \cite{TehJor2010a}, we use another 
representation that requires no dynamic memory and has better inference 
efficiency \citep{Chen:2011:STC:2034063.2034095}. 

We denote $f^*(\mathcal{N})$ as the marginalised likelihood associated 
with the probability vector $\mathcal{N}$. Since the vector is 
marginalised out, the marginalised likelihood is in terms of --- 
using the CRP terminology --- the \textit{customer counts} 
$c^\mathcal{N} = (\cdots, c_k^\mathcal{N}, \cdots)$ and the 
\textit{table counts} $t^\mathcal{N} = (\cdots, t_k^\mathcal{N}, 
\cdots)$. The customer count $c_k^\mathcal{N}$ corresponds to the 
number of data points (\textit{e.g.}\ words) assigned to group $k$ 
(\textit{e.g.}\ topic) for variable $\mathcal{N}$. Here, the 
\textit{table counts} $t^\mathcal{N}$ represent the subset of 
$c^\mathcal{N}$ that gets passed up the hierarchy (as customers for 
the parent probability vector of $\mathcal{N}$). Thus 
$t_k^\mathcal{N} \leq c_k^\mathcal{N}$, and $t_k^\mathcal{N}=0$ if and 
only if $c_k^\mathcal{N}=0$ since the counts are non-negative.
We also denote $C^\mathcal{N} = \sum_k c_k^\mathcal{N}$ as the total 
customer counts for node $\mathcal{N}$, and similarly, 
${T}^\mathcal{N} = \sum_k t_k^\mathcal{N}$ is the total table counts. 
The marginalised likelihood $f^*(\mathcal{N})$, in terms of 
$c^\mathcal{N}$ and $t^\mathcal{N}$, is given as
\begin{align}
f^*(\mathcal{N}) & = 
\frac{(\beta^\mathcal{N}|\alpha^\mathcal{N})_{{T}^\mathcal{N}}}
{(\beta^\mathcal{N})_{C^\mathcal{N}}} 
\prod_k 
S^{c^\mathcal{N}_k}_{{t}^\mathcal{N}_k, \alpha^\mathcal{N}} ~~,
&&
\mathrm{for\ \ }
\mathcal{N} \sim 
\mathrm{PYP}(\alpha^\mathcal{N}, \beta^\mathcal{N}, \mathcal{P}) 
~.  
\label{eq:modularized_likelihood} 
\end{align} 
$S^x_{y,\alpha}$ is the generalised Stirling number that is easily 
tabulated; both $(x)_C$ and $(x|y)_C$ denote the Pochhammer symbol 
(rising factorial), see \cite{2012arXiv1007.0296B} for details. Note the 
GEM distribution behaves like a PYP in which the table count 
$t_k^\mathcal{N}$ is always $1$ for non-zero $c_k^\mathcal{N}$.

The innovation of \cite{Chen:2011:STC:2034063.2034095} was to notice 
that sampling with \Cref{eq:modularized_likelihood} directly 
led to poor performance.  The problem was that sampling an assignment
to a latent variable, say moving a customer from group $k$ to $k'$
(so $c_k^\mathcal{N}$ decreases by $1$ and $c_{k'}^\mathcal{N}$ 
increases by $1$), the potential effect on $t_k^\mathcal{N}$ and  
$t_{k'}^\mathcal{N}$ could not immediately be measured. Whereas, the 
hierarchical CRP automatically included table configurations in its 
sampling process and thus included the influence of the hierarchy in the 
sampling. Thus sampling directly with 
\Cref{eq:modularized_likelihood} lead to comparatively poor 
mixing. As a solution, \cite{Chen:2011:STC:2034063.2034095} develop a 
collapsed version of the hierarchical CRP following the well known 
practice of Rao-Blackwellisation of sampling schemes 
\citep{casella1996rao}, which, while not being as fast per step, it has 
two distinct advantages, (1) it requires no dynamic memory and (2) the 
sampling has significantly lower variance so converges much faster. This 
has empirically been shown to lead to better mixing of the samplers 
\citep{Chen:2011:STC:2034063.2034095} and has been confirmed on 
different complex topic models \citep{Buntine:2014:ENT:2623330.2623691}.

The technique for collapsing the  hierarchical CRP uses 
\Cref{eq:modularized_likelihood} but the counts
($c^\mathcal{N},t^\mathcal{N}$) are now derived variables.
They are derived from Boolean variables associated with each data point.
The technique comprises the following conceptual steps:
(1) add Boolean indicators $u_{dn}$ to the data $(z_{dn},w_{dn})$ from 
which the counts $c^\mathcal{N}$ and $t^\mathcal{N}$ can be derived,
(2) modify the marginalised posterior accordingly, and
(3) derive a sampler for the model.

\subsubsection{Adding Boolean indicators}
We first consider $c_k^{\theta_d}$, which has a ``+1'' contributed to 
for every $z_{dn}=k$ in document $d$, hence 
$c_k^{\theta_d}=\sum_n I(z_{dn}=k)$. We now introduce a new Bernoulli 
\textit{indicator variable} $u^{\theta_d}_{dn}$ associated with 
$z_{dn}$, which is ``on'' (or $1$) when the data $z_{dn}$ also 
contributed a ``+1'' to  $t^{\theta_d}_k$. Note that 
$t_k^{\theta_d} \leq c_k^{\theta_d}$, so every data contributing a 
``+1'' to $c_k^{\theta_d}$ may or may not contribute a ``+1'' to 
$t_k^{\theta_d}$. The result is that one derives 
$t_k^{\theta_d}=\sum_n I(z_{dn}=k) \, I(u^{\theta_d}_{dn}=1)$.

Now consider the parent of $\theta_d$, which is $\theta'_d$.
Its customer count is derived as $c_k^{\theta'_d}=t_k^{\theta_d}$.
Its table count $t_k^{\theta'_d}$ can now be treated similarly.
Those data $z_{dn}$ that contribute a ``+1'' to $t_k^{\theta_d}$
(and thus $c_k^{\theta'_d}$) have a new Bernoulli indicator variable 
$u^{\theta'_d}_{dn}$, which is used to derive 
$t_k^{\theta'_d}=\sum_n I(z_{dn}=k) \, I(u^{\theta'_d}_{dn}=1)$,
similar as before. Note that if $u^{\theta'_d}_{dn}=1$ then necessarily 
$u^{\theta_d}_{dn}=1$.

Similarly, one can define Boolean indicators for $\mu$, $\nu_b$, 
$\phi'$, $\phi$, and $\gamma$ to have a full suite from which all the 
counts $c^\mathcal{N}$ and $t^\mathcal{N}$ are now derived. We denote 
$u_{dn} = \{ 
u^{\theta_d}_{dn}, u^{\theta'_d}_{dn}, u^{\nu_b}_{dn}, u^{\mu}_{dn}, 
u^{\phi'_d}_{dn}$, $u^{\phi_d}_{dn}, u^\gamma_{dn} \}$ 
as the collection of the Boolean indicators for data ($z_{dn}$, $w_{dn}$).

\subsubsection{Probability of Boolean indicators}

By symmetry, if there are $t_k^\mathcal{N}$ Boolean indicators ``on'' 
(out of $c_k^\mathcal{N}$), we are indifferent as to which is on. Thus 
the indicator variable $u^\mathcal{N}_{dn}$ is not stored, that is, we 
simply ``forget'' who contributed a table count and re-sample 
$u^\mathcal{N}_{dn}$ as needed:
\begin{align}
p(u^\mathcal{N}_{dn} = 1) = t^{\mathcal{N}}_k / c^\mathcal{N}_k ~~, &&
p(u^\mathcal{N}_{dn} = 0) = 1 - t^{\mathcal{N}}_k / c^\mathcal{N}_k ~~.
\label{eq:sample_u}
\end{align}
Moreover, this means that the marginalised 
likelihood $f^*(\mathcal{N})$ of \Cref{eq:modularized_likelihood}
is extended to include the probability of $u^\mathcal{N}$, which
is written in terms of $c^\mathcal{N}$, $t^\mathcal{N}$ and 
$u^\mathcal{N}$ as:
\begin{equation}
f(\mathcal{N})  = f^*(\mathcal{N}) \,
 p\big( u^\mathcal{N}  \,\big|\,  c^\mathcal{N}, {t}^\mathcal{N} \big) 
 = f^*(\mathcal{N})
\prod_k {\binom{c^\mathcal{N}_k}{{t}^\mathcal{N}_k}}^{-1}
~~~.  
\label{eq:modularized_likelihood_boolean} 
\end{equation}

\subsection{Likelihood for the Hierarchical PYP Topic Model}

We use bold face capital letters to denote the set of all relevant 
lower case variables.  For example, 
$\mathbf{Z} = \{z_{11},\cdots,z_{DN_D}\}$ denotes the set of all topic 
assignments. Variables $\mathbf{W}$, $\mathbf{T}$, $\mathbf{C}$ and 
$\mathbf{U}$ are similarly defined, that is, they denote the set of all 
words, table counts, customer counts, and Boolean indicators 
respectively. Additionally, we denote $\mathbf{\zeta}$ as the set of all 
hyperparameters (such as the $\alpha$'s). With the probability vectors 
replaced by the counts, the likelihood of the topic model can be written 
--- in terms of $f(\cdot)$ as given in 
\Cref{eq:modularized_likelihood_boolean} --- as 
$p(\mathbf{Z}, \mathbf{W}, \mathbf{T}, \mathbf{C}, \mathbf{U} 
 \,|\,  \mathbf{\zeta}) \propto$
\begin{align}
f(\mu) \Bigg(\prod_{b=1}^B f(\nu_b) \! \Bigg) 
\Bigg(\prod_{d=1}^D f(\theta'_d) \,
f(\theta_d)
\prod_{k=1}^K f(\phi'_{dk}) \! \Bigg)
\Bigg(\prod_{k=1}^K f(\phi_k) \! \Bigg)
f(\gamma) 
\Bigg( \! 
\prod_v \left(\frac{1}{|\mathcal{V}|}\right)^{t^\gamma_v} 
\Bigg) ~.
\label{eq:likelihood}
\end{align}
Note that the last term in \Cref{eq:likelihood} corresponds 
to the parent probability vector of $\gamma$ (see 
\Cref{subsec:topic_model}), and $v$ indexes the unique word 
tokens in vocabulary set $\mathcal{V}$. Note that the extra terms for 
$\mathbf{U}$ are simply derived using 
\Cref{eq:modularized_likelihood_boolean} and not
stored in the model. So in the discussions below we will usually 
represent $\mathbf{U}$ implicitly by $\mathbf{T}$ and $\mathbf{C}$,
and introduce the $\mathbf{U}$ when explicitly needed.

Note that even though the probability vectors are integrated out and 
not explicitly stored, they can easily be estimated from the associated 
counts. The probability vector $\mathcal{N}$ can be estimated from its 
posterior mean given the counts and parent probability vector 
$\mathcal{P}$:
\begin{align}
\hat{\mathcal{N}} &= 
\left( \cdots , 
\frac{
(\alpha^\mathcal{N} T^\mathcal{N} + \beta^\mathcal{N}) \mathcal{P}_k 
+ c_k^\mathcal{N} - \alpha^\mathcal{N} T_k^\mathcal{N}
}{
\beta^\mathcal{N} + C^\mathcal{N}
} 
, \cdots \right) ~.
\label{eq:recover_vector}
\end{align}

\subsection{Likelihood for the Citation Network Poisson Model}

For the citation network, the Poisson likelihood for each $x_{ij}$ 
is given as
\begin{align}
p(x_{ij} \,|\, \lambda, \theta) = 
\frac{\lambda_{ij}^{x_{ij}}}{x_{ij}! \, e^{\lambda_{ij}}} 
\approx \left( 
\lambda_i^+ \lambda_j^- 
\sum_k \lambda^T_k \theta'_{ik} \theta'_{jk} 
\right)^{x_{ij}}
\exp \left(- \lambda_i^+ \lambda_j^-  
\sum_k \lambda^T_k \theta'_{ik} \theta'_{jk} \right)
~.
\label{eq:poisson_likelihood}
\end{align}
Note that the term $x_{ij}!$ is dropped in 
\Cref{eq:poisson_likelihood} due to the limitation of the data 
that $x_{ij} \in \{0, 1\}$, thus $x_{ij}!$ is evaluated to $1$. With 
conditional independence of $x_{ij}$, the joint likelihood for the whole 
citation network $\mathbf{X} = \{x_{11}, \cdots, x_{DD}\}$ can be 
written as $p(\mathbf{X} \,|\, \lambda, \theta') =$
\begin{align}
 \left( \prod_i 
 	(\lambda_i^{+})^{g^+_i} \, (\lambda_i^{-})^{g^-_i} 
 \right)
 \prod_{ij} 
 \left( 
 	\sum_k \lambda^T_k \theta'_{ik} \theta'_{jk} 
 \right)^{\! x_{ij}} 
 \exp\!\Bigg( 
 	- \sum_{ijk} \lambda_i^+ \lambda_j^- \lambda^T_k 
 	\theta'_{ik} \theta'_{jk} 
 \Bigg) 
 ~,
 \label{eq:citation_likelihood}
\end{align}
where $g^+_i$ is the number of citations for publication $i$, 
$g^+_i = \sum_j x_{ij}$, and $g^-_i$ is the number of times 
publication $i$ being cited, $g^-_i = \sum_j x_{ji}$. We also make a 
simplifying assumption that $x_{ii} = 1$ for all documents $i$, that is, 
all publications are treated as self-cited. This assumption is important 
since defining $x_{ii}$ allows us to rewrite the joint likelihood into 
\Cref{eq:citation_likelihood},
which leads to a cleaner learning algorithm that utilises 
an efficient caching.
Note that if we do not define $x_{ii}$, we have to explicitly consider
the case when $i=j$ in \Cref{eq:citation_likelihood}
which results in messier summation and products. 

Note the likelihood in \Cref{eq:citation_likelihood} contains
the document-topic distribution $\theta'$ in vector form. This is 
problematic as performing inference with the likelihood requires the 
probability vectors $\theta'$, $\nu$ and $\mu$ to be stored explicitly 
(instead of counts as discussed in \Cref{subsec:mhpyp}).
To overcome this issue, we propose a novel representation that allows 
the probability vectors to remain integrated out. Such representation 
also leads to an efficient sampling algorithm for the citation network, 
as we will see in \Cref{sec:inference}.

We introduce an \textit{auxiliary variable} $y_{ij}$, named the
\textit{citing topic}, to denote the topic that prompts publication 
$i$ to cite publication $j$. To illustrate, for a \textit{biology} 
publication that cites a \textit{machine learning} publication for the 
learning technique, the citing topic would be `machine learning' instead 
of `biology'. From \Cref{eq:lambda_ij}, we model the citing 
topic $y_{ij}$ as jointly Poisson with $x_{ij}$:
\begin{align}
x_{ij}, y_{ij} = k  \,|\,  \lambda, \theta'
\sim \mathrm{Poisson}
 \left(\lambda^+_i \lambda^-_j \lambda^T_k \theta'_{ik} \theta'_{jk} \right)~.
\label{eq:citing_topic}
\end{align}
Incorporating $\mathbf{Y}$, the set of all $y_{ij}$, 
we rewrite the citation network likelihood as
\begin{align}
p(\mathbf{X},\mathbf{Y}|\lambda, \theta') \propto
\prod_i (\lambda_i^+)^{g^+_i} (\lambda_i^-)^{g^-_i} 
\prod_k \left(\lambda_k^T\right)^{\frac{1}{2}\sum_i h_{ik}} \!
\prod_{ik} {\theta'_{ik}}^{h_{ik}}
\, \exp \! \Bigg(- 
\sum_{ij} \lambda_i^+ \lambda_j^-  \lambda^T_{y_{ij}} 
\theta'_{i{y_{ij}}} \theta'_{j{y_{ij}}} \Bigg) ~,
\end{align}
where $h_{ik}=\sum_j x_{ij}I(y_{ij}=k)+\sum_j x_{ji}I(y_{ji}=k)$ 
is the number of connections publication $i$ made due to topic $k$.

To integrate out $\theta'$, we note the term ${\theta'_{ik}}^{h_{ik}}$ 
appears like a multinomial likelihood, so we absorb them into the 
likelihood for 
$p(\mathbf{Z}, \mathbf{W}, \mathbf{T}, \mathbf{C}, \mathbf{U} \,|\, \mathbf{\zeta})$ 
where they correspond to additional counts for $c^{\theta'_i}$, with 
$h_{ik}$ added to $c^{\theta'_i}_k$. To disambiguate the source of 
the counts, we will refer to these customer counts contributed by 
$x_{ij}$ as \textit{network counts}, and denote the augmented counts 
($\mathbf{C}$ plus network counts) as $\mathbf{C^+}$.
For the exponential term, we use the delta method \citep{delta_method}
to approximate
$\int q(\theta)\,\exp(-g(\theta))\,\mathrm{d}\theta
\approx \exp(-g(\hat\theta)) \int q(\theta)\,\mathrm{d}\theta$,
where $\hat\theta$ is the expected value according to a distribution
proportional to $q(\theta)$. This approximation is reasonable as 
long as the terms in the exponential are small 
(see \Cref{appendix:delta_method}). The approximate full 
posterior of SCNTM can then be written as 
$p(\mathbf{Z}, \mathbf{W}, \mathbf{T}, \mathbf{C^+}, \mathbf{U}, 
\mathbf{X}, \mathbf{Y} \,|\, \lambda,\mathbf{\zeta}) \approx$
\begin{align}
\!\! p(\mathbf{Z}, \mathbf{W}, \mathbf{T}, \mathbf{C^+}, \mathbf{U}  \,|\,  \mathbf{\zeta})
\prod_i (\lambda_i^+)^{g^+_i} (\lambda_i^-)^{g^-_i} 
\prod_k (\lambda_k^T)^{g_k^T}
\, \exp \! 
\Bigg( \!\!\! - \sum_{ij} \lambda_i^+ \lambda_j^-  \lambda^T_{y_{ij}} 
\hat\theta'_{iy_{ij}} \hat\theta'_{jy_{ij}} \!\!
\Bigg) ,
\label{eq:approx_network_likelihood}
\end{align}
where $g_k^T = \frac{1}{2}\sum_i h_{ik}$\,. 
We note that 
$p(\mathbf{Z}, \mathbf{W}, \mathbf{T}, \mathbf{C^+}, \mathbf{U} 
 \,|\,  \mathbf{\zeta})$ is the same as \Cref{eq:likelihood} but now 
with $\mathbf{C^+}$ instead of $\mathbf{C}$.

In the next section, we demonstrate that our model representation 
gives rise to an intuitive sampling algorithm for learning the model.
We also show how the Poisson model integrates into the topic 
modelling framework.

\section{Inference Techniques}
\label{sec:inference}

Here, we derive the Markov chain Monte Carlo (MCMC) algorithms for 
learning the SCNTM. We first describe the sampler for the topic model 
and then for the citation network. The full inference procedure is 
performed by alternating between the two samplers. Finally, we outline 
the hyperparameter samplers that are used to estimate the 
hyperparameters automatically.

\subsection{Sampling for the Hierarchical PYP Topic Model}
\label{subsec:gibbs_sampler}

To sample the words' topic $\mathbf{Z}$ and the associated counts 
$\mathbf{T}$ and $\mathbf{C}$ in the SCNTM, we design a 
Metropolis-Hastings (MH) algorithm based on the collapsed Gibbs sampler 
designed for the PYP \citep{Chen:2011:STC:2034063.2034095}. The concept 
of the MH sampler is analogous to LDA, which consists of 
(1) decrementing the counts associated with a word, 
(2) sampling the respective new topic assignment for the word, and 
(3) incrementing the associated counts. 
However, our sampler is more complicated than LDA. In particular, we 
have to consider the indicators $u^\mathcal{N}_{dn}$ described in 
\Cref{subsec:mhpyp} operating on the hierarchy of PYPs.
Our MH sampler consists of two steps. First we sample the latent topic
$z_{dn}$ associated with the word $w_{dn}$. We then sample the customer 
counts $\mathbf{C}$ and table counts $\mathbf{T}$.

The sampler proceeds by considering the latent variables associated 
with a given word $w_{dn}$. First, we decrement the counts associated 
with the word $w_{dn}$ and the latent topic $z_{dn}$. This is achieved 
by sampling the suite of indicators $u_{dn}$ according to 
\Cref{eq:sample_u} and decrementing the relevant customer counts 
and table counts. For example, we decrement $c^{\theta_d}_{z_{dn}}$ by 1
if $u^{\theta_d}_{dn} = 1$. After decrementing, we apply a Gibbs sampler 
to sample a new topic $z_{dn}$ from its conditional posterior 
distribution, given as
\begin{align}
& p(z^\mathrm{new}_{dn}  \,|\,  \mathbf{Z}^{-dn}, \mathbf{W}, 
\mathbf{T}^{-dn}, \mathbf{C^+}^{-dn}, \mathbf{U}^{-dn}, \mathbf{\zeta}) 
\nonumber
\\
& =
\sum_{u_{dn}} p \big(z^\mathrm{new}_{dn}, u_{dn}  \,\big|\,  \mathbf{Z}^{-dn}, \mathbf{W}, 
\mathbf{T}^{-dn}, \mathbf{C^+}^{-dn}, \mathbf{U}^{-dn}, \mathbf{\zeta} \big) 
\,. 
\label{eq:conditional_posterior}
\end{align}
Note that the joint distribution in \Cref{eq:conditional_posterior} can be 
written as the ratio of the likelihood for the topic model 
(\Cref{eq:likelihood}):
\begin{align}
\frac{
p(\mathbf{Z}, \mathbf{W}, \mathbf{T}, \mathbf{C^+}, \mathbf{U}  \,|\,  \mathbf{\zeta}) 
}{
p(\mathbf{Z}^{-dn}, \mathbf{W}, \mathbf{T}^{-dn}, \mathbf{C^+}^{-dn}, 
\mathbf{U}^{-dn}  \,|\,  \mathbf{\zeta})
} 
~.
\end{align}
Here, the superscript $\Box^{-dn}$ indicates that the topic $z_{dn}$, indicators 
and the associated counts for word $w_{dn}$ are not observed in the respective 
sets, \textit{i.e.}\ the state after decrement. Additionally,  we use the 
superscripts $\Box^\mathrm{new}$ and $\Box^\mathrm{old}$ to denote the proposed 
sample and the old value respectively. The modularised likelihood of 
\Cref{eq:likelihood} allows the conditional posterior 
(\Cref{eq:conditional_posterior}) to be computed easily, since it 
simplifies to ratios of likelihood $f(\cdot)$, which simplifies further since the 
counts differ by at most $1$ during sampling. For instance, the ratio of the 
Pochhammer symbols, $(x|y)_{C+1} / (x|y)_C$, simplifies to  $x+Cy$, while the 
ratio of Stirling numbers, such as $S^{y+1}_{x+1, \alpha}/S^{y}_{x, \alpha}$, can 
be computed quickly via caching~\citep{2012arXiv1007.0296B}.

Next, we proceed to sample the relevant customer counts and table counts given
the new $z_{dn} = k$. We propose an MH algorithm for this. We define the proposal 
distribution for the new customer counts and table counts as
\begin{align}
q \Big(\mathbf{T}^\mathrm{new}, \mathbf{C^+}^\mathrm{new}  \,\Big|\,  \mathbf{Z}, \mathbf{W}, 
\mathbf{T}^{-dn}, \mathbf{C^+}^{-dn}, \mathbf{\zeta} \Big)
& \propto
\frac{ 
p \big(\mathbf{Z}, \mathbf{W}, \mathbf{T}^\mathrm{new}, \mathbf{C^+}^\mathrm{new}, \mathbf{U}^\mathrm{new} 
   \,\big|\,  \mathbf{\zeta} \big)
}{
p \big(\mathbf{Z}, \mathbf{W}, \mathbf{T}^{-dn}, \mathbf{C^+}^{-dn}, \mathbf{U}^{-dn}  \,\big|\,  \mathbf{\zeta} \big)
}
\label{eq:sample_counts}
\end{align}
where 
\begin{align}
p(\mathbf{Z}, \mathbf{W}, \mathbf{T}, \mathbf{C^+}, \mathbf{U}  \,|\,  \mathbf{\zeta})
& \propto
f(\mu) \Bigg(\prod_{b=1}^B f(\nu_b) \! \Bigg) 
\Bigg(\prod_{d=1}^D f(\theta'_d) \,
f(\theta_d)
\prod_{k=1}^K f(\phi'_{dk}) \! \Bigg)
\nonumber \\
& \hspace{5mm}
\Bigg(\prod_{k=1}^K f(\phi_k) \! \Bigg)
f(\gamma) 
\Bigg( \! 
\prod_v \left(\frac{1}{|\mathcal{V}|}\right)^{t^\gamma_v} 
\Bigg) ~.
\end{align}
Here, the potential sample space for $\mathbf{T}^\mathrm{new}$ and 
$\mathbf{C}^\mathrm{new}$ are restricted to just $t_k + i$ and $c_k + i$
where $i$ is either $0$ or $1$. Doing so allows us to avoid considering the 
exponentially many possibilities of $\mathbf{T}$ and $\mathbf{C}$. The acceptance 
probability associated with the newly sampled $\mathbf{T}^\mathrm{new}$ and 
$\mathbf{C}^\mathrm{new}$ is
\begin{align}
A & =
\frac{ 
p \big(\mathbf{Z}, \mathbf{W}, \mathbf{T}^\mathrm{new}, \mathbf{C^+}^\mathrm{new}, 
\mathbf{U}^\mathrm{new}  \,\big|\,  \mathbf{\zeta} \big)
}{
p \big(\mathbf{Z}, \mathbf{W}, \mathbf{T}^{old}, \mathbf{C^+}^{old}, \mathbf{U}^{old}  
\,\big|\,  \mathbf{\zeta} \big)
}
\cdot
\frac{
q\big(\mathbf{T}^\mathrm{old}, \mathbf{C^+}^\mathrm{old}  \,\big|\,  \mathbf{Z}, \mathbf{W}, \mathbf{T}^{-dn}, \mathbf{C^+}^{-dn} \mathbf{\zeta} \big)
}{
q\big(\mathbf{T}^\mathrm{new}, \mathbf{C^+}^\mathrm{new}  \,\big|\,  \mathbf{Z}, \mathbf{W}, \mathbf{T}^{-dn}, \mathbf{C^+}^{-dn} \mathbf{\zeta} \big)
}
\nonumber \\
& = 1 ~~~.
\end{align}
Thus we always accept the proposed sample.%
\footnote{The algorithm is named MH algorithm instead of Gibbs sampling 
due to the fact that the sample space for the counts is restricted and
thus we are not sampling from the posterior directly.}
Note that since $\mu$ is GEM distributed, incrementing $t^\mu_k$ is equivalent to 
sampling a \textit{new} topic, \textit{i.e.}\ the number of topics increases by 
$1$.

\subsection{Sampling for the Citation Network}

For the citation network, we propose another MH algorithm. The MH algorithm can be 
summarised in three steps: 
(1) estimate the document topic prior $\theta'$, 
(2) propose a new citing topic $y_{ij}$, and 
(3) accept or reject the proposed $y_{ij}$ following an MH scheme.
Note that the MH algorithm is similar to the sampler for the topic model, 
where we decrement the counts, sample a new state and update the counts. Since all 
probability vectors are represented as counts, we do not need to deal with their 
vector form. Additionally, our MH algorithm is intuitive and simple to implement. 
Like the words in a document, each citation is assigned a topic, hence the words 
and citations can be thought as voting to determine a documents' topic. 

We describe our MH algorithm for the citation network as follows. 
First, for each document~$d$, we estimate the expected document-topic prior 
$\hat\theta'_d$ from \Cref{eq:recover_vector}. Then, for each document 
pair $(i, j)$ where $x_{ij}=1$, we decrement the network counts associated with 
$x_{ij}$, and re-sample $y_{ij}$ with a proposal distribution derived from 
\Cref{eq:citing_topic}:
\begin{align}
p(y^\mathrm{new}_{ij} = k  \,|\,  \hat\theta'_i, \hat\theta'_j) 
\propto 
\lambda^T_k \hat\theta'_{ik} \hat\theta'_{jk} \, 
\exp \left( - \lambda^+_i \lambda^-_j \lambda^T_k \hat\theta'_{ik} \hat\theta'_{jk} \right) ~~~,
\end{align}
which can be further simplified since the terms inside the exponential are very 
small, hence the exp term approximates to $1$. We empirically inspected the 
exponential term and we found that almost all of them are between $0.99$ and $1$. 
This means the ratio of the exponentials is not significant for sampling new 
citing topic $y_{ij}^\mathrm{new}$. So we ignore the exponential term and let
\begin{align}
p(y^\mathrm{new}_{ij} = k  \,|\,  \hat\theta'_i, \hat\theta'_j) 
\propto 
\lambda^T_k \hat\theta'_{ik} \hat\theta'_{jk} ~~~.
\label{eq:proposal_y}
\end{align}
We compute the acceptance probability $A$ for the newly sampled 
$y_{ij}^\mathrm{new}=y'$, changed from $y^\mathrm{old}_{ij}=y^*$, and the 
successive change to the document-topic priors (from 
$\mbox{$\hat\theta'$}^\mathrm{old}$ to $\mbox{$\hat\theta'$}^\mathrm{new}$):
\begin{align}
A \ = \
& \frac{
\exp\left(- \sum_{ijk} \lambda_i^+ \lambda_j^-\lambda_k^T 
\mbox{${\hat\theta'}_{ik}$}^{\mathrm{\!\!new}} 
\mbox{$\hat\theta'_{jk}$}^{\mathrm{\!\!new}} \right)
}{
\exp\left(- \sum_{ijk} \lambda_i^+ \lambda_j^- \lambda_k^T 
\mbox{$\hat\theta'_{ik}$}^{\mathrm{\!\!old}} 
\mbox{$\hat\theta'_{jk}$}^{\mathrm{\!\!old}} \right)
}
\frac{
p(\mathbf{Z}, \mathbf{W}, \mathbf{T}, \mathbf{C^+}^{\mathrm{new}}, \mathbf{U}  \,|\,  \mathbf{\zeta})
}{
p(\mathbf{Z}, \mathbf{W}, \mathbf{T}, \mathbf{C^+}^{\mathrm{old}}, \mathbf{U}  \,|\,  \mathbf{\zeta})
}
\times
\nonumber
\\
& \frac{
\lambda^T_{y^*} \mbox{$\hat\theta'_{i{y^*}}$}^{\mathrm{\!\!\!new}} 
\mbox{$\hat\theta'_{j{y^*}}$}^{\mathrm{\!\!\!new}}
}{
\lambda^T_{y'} \mbox{$\theta'_{iy'}$}^{\mathrm{\!\!\!old}} 
\mbox{$\theta'_{jy'}$}^{\mathrm{\!\!\!old}} 
} 
\frac{
\sum_k \lambda^T_{k} \mbox{$\hat\theta'_{ik}$}^{\mathrm{\!\!old}} 
\mbox{$\hat\theta'_{jk}$}^{\mathrm{\!\!old}}
}{
\sum_k \lambda^T_k \mbox{$\hat\theta'_{ik}$}^{\mathrm{\!\!new}} 
\mbox{$\hat\theta'_{jk}$}^{\mathrm{\!\!new}}
}
~~~.
\label{eq:acceptance_probability}
\end{align}
Note that we have abused the notations $i$ and $j$ in the above equation,
where the $i$ and $j$ in the summation indexes all documents instead of pointing 
to particular document $i$ and document $j$. We decided against introducing 
additional variables to make things less confusing.

Finally, if the sample is accepted, we update $y_{ij}$ and the associated customer 
counts. Otherwise, we discard the sample and revert the changes.

\subsection{Hyperparameter Sampling}
\label{subsec:hyperparameter_sampling}

Hyperparameter sampling for the priors are important~\citep{WallachPrior2009}.
In our inference algorithm, we sample the concentration parameters $\beta$ of all 
PYPs with an auxiliary variable sampler \citep{Teh06abayesian}, but leave the 
discount parameters $\alpha$ fixed. We do not sample the $\alpha$ due to the 
coupling of the parameter with the Stirling numbers cache.

Here we outline the procedure to sample the concentration parameter 
$\beta^\mathcal{N}$ of a PYP distributed variable $\mathcal{N}$, using an 
auxiliary variable sampler. Assuming each $\beta^\mathcal{N}$ has a Gamma 
distributed hyperprior with shape $\tau_0$ and rate $\tau_1$, we first sample the 
auxiliary variables $\xi$ and $\psi_j$ for $j \in \{0, T^\mathcal{N} -1 \}$:
\begin{align}
\xi  \,|\,  \beta^\mathcal{N} \sim \mathrm{Beta}(C^\mathcal{N}, \beta^\mathcal{N}) ~~,
&&
\psi_j  \,|\,  \alpha^\mathcal{N}, \beta^\mathcal{N} \sim 
\mathrm{Bernoulli}\left(\frac{\beta^\mathcal{N}}{\beta^\mathcal{N}+j\alpha^\mathcal{N}}\right) ~~.
\end{align}
We then sample a new $\beta'{^\mathcal{N}}$ from the following conditional 
posterior given the auxiliary variables:
\begin{align}
\beta'{^\mathcal{N}}  \,|\,  \xi, \psi \sim \mathrm{Gamma}\left(\tau_0 + \textstyle{ \sum_j } \psi_j, \tau_1 - \log(1 - \xi)\right) ~~~.
\end{align}

In addition to the PYP hyperparameters, we also sample $\lambda^+$, $\lambda^-$ 
and $\lambda^T$ with a Gibbs sampler. We let the hyperpriors for $\lambda^+$, 
$\lambda^-$ and $\lambda^T$ to be Gamma distributed with shape $\epsilon_0$ and 
rate $\epsilon_1$. With the conjugate Gamma prior, the posteriors for 
$\lambda^+_i$, $\lambda^-_i$ and $\lambda^T_k$ are also Gamma distributed, so they 
can be sampled directly.
\begin{align}
(\lambda^+_i \,|\, \mathbf{X}, \lambda^-, \lambda^T \theta') 
& \sim 
\textstyle
\mathrm{Gamma}\left( \epsilon_0 + g^+_i, 
\epsilon_1 + \sum_k \lambda^T_k \theta'_{ik} 
\sum_j \lambda^-_j \theta'_{jk} \right) ~~~,
\\
(\lambda^-_i \,|\, \mathbf{X}, \lambda^+, \lambda^T \theta') 
& \sim 
\textstyle
\mathrm{Gamma}\left( \epsilon_0 + g^-_i, 
\epsilon_1 + \sum_k \lambda^T_k \theta'_{ik} 
\sum_j \lambda^+_j \theta'_{jk} \right) ~~~,
\\
(\lambda^T_k \,|\, \mathbf{X}, \mathbf{Y}, \lambda^+, \lambda^-, \theta') 
& \sim 
\textstyle
\mathrm{Gamma}\left( \epsilon_0 + \frac{1}{2} \sum_i h_{ik}, 
\epsilon_1 + \lambda^T_k (\sum_j \lambda^+_j \theta'_{jk}) 
(\sum_j \lambda^-_j \theta'_{jk}) \right) .
\end{align}
We apply vague priors to the hyperpriors by setting 
$\tau_0 = \tau_1 = \epsilon_0 = \epsilon_1 = 1$.

\begin{algorithm}[t!]
	\caption{Inference Algorithm for the Citation Network Topic Model}
	\label{alg:gibbs}
	\begin{enumerate}[itemindent=0pt, itemsep=1pt]
		\item 
			Initialise the model by assigning a random topic 
			assignment $z_{dn}$ to each word $w_{dn}$ and 
			constructing the relevant customer counts 
			$c^\mathcal{N}_k$ and table counts $t^\mathcal{N}_k$ for 
			all variables $\mathcal{N}$.
		\item 
			For each word $w_{dn}$ in each document $d$:
			\begin{enumerate}[itemindent=0pt, noitemsep, nolistsep,
				label=\roman{*}., ref=(\roman{*})]
  				\item 
  					Decrement the counts associated with $z_{dn}$ and 
  					$w_{dn}$~.
		    	\item 
		    		Sample a new topic $z_{dn}$ with
		    		its conditional posterior in 
		    		\Cref{eq:conditional_posterior}.
		    	\item
		    		Sample the counts $\mathbf{T}$ and $\mathbf{C}$
		    		with the proposal distribution in \Cref{eq:sample_counts}.
			\end{enumerate}
		\item 
			For each citation $x_{ij} = 1$:
			\begin{enumerate}[itemindent=0pt, noitemsep, nolistsep, 
				label=\roman{*}., ref=(\roman{*})]
  				\item 
  					Decrement the network counts associated with 
		  			$x_{ij}$ and $y_{ij}$~.
			    \item 
			    	Sample a new citing topic $y_{ij}$ with the proposal
			    	distribution in \Cref{eq:proposal_y}.
		   	 	\item 
		   	 		Accept or reject the sampled $y_{ij}$ with the acceptance 
				    probability in \Cref{eq:acceptance_probability}.
			\end{enumerate}
		\item 
			Update the hyperparameters $\beta$, $\lambda^+$,
			$\lambda^-$ and $\lambda^T$.
		\item 
			Repeat steps 2-4 
			until the model converges or a fix number of iterations 
			reached.
	\end{enumerate}
\end{algorithm}

Before we proceed with the next section on the datasets used in the paper,
we summarise the full inference algorithm for the SCNTM in 
\Cref{alg:gibbs}.

\section{Data}
\label{sec:data}

We perform our experiments on subsets of CiteSeer$^\mathrm{X}$ 
data\footnote{\url{http://citeseerx.ist.psu.edu/}} which consists of 
scientific publications. Each publication from CiteSeer$^\mathrm{X}$ 
is accompanied by {\it title}, {\it abstract}, {\it keywords}, 
{\it authors}, {\it citations} and other metadata. We prepare three 
publication datasets from CiteSeer$^\mathrm{X}$ for evaluations.
The first dataset corresponds to Machine Learning (ML) publications, 
which are queried from CiteSeer$^\mathrm{X}$ using the keywords from 
Microsoft Academic 
Search.\footnote{\url{http://academic.research.microsoft.com/}}
The ML dataset contains 139,227 publications.
Our second dataset corresponds to publications from ten distinct 
research areas.
The query words for these ten disciplines are chosen such that the 
publications form distinct clusters. We name this dataset M10 
(Multidisciplinary 10 classes), which is made of 10,310 publications.
For the third dataset, we query publications from both arts and 
science disciplines. Arts publications are made of \textit{history} 
and \textit{religion} publications, while the science publications 
contain \textit{physics}, \textit{chemistry} and \textit{biology} 
research. This dataset consists of 18,720 publications and is 
named AvS (Arts versus Science) in this paper.
These queried datasets are made available online.%
\footnote{\url{http://karwai.weebly.com/publications.html}}

The keywords used to create the datasets are obtained from Microsoft 
Academic Search, and are listed in \Cref{appendix:keywords}.
For the clustering evaluation in \Cref{subsubsec:clustering}, 
we treat the query categories as the ground truth. However, 
publications that span multiple disciplines can be problematic for 
clustering evaluation, hence we simply remove the publications that 
satisfy the queries from more than one discipline. Nonetheless, the 
labels are inherently noisy. The metadata for the publications can 
also be noisy, for instance, the {\it authors} field may sometimes 
display publication's keywords instead of the authors, publication 
title is sometimes an URL, and table of contents can be mistakenly 
parsed as the abstract. We discuss our treatments to these issues in 
\Cref{subsec:preprocessing}. We also note that non-English 
publications are discarded using 
{\tt langid.py}~\citep{Lui:2012:LOL:2390470.2390475}.

\begin{table}[t]
	\centering
	\caption{
		Summary of the datasets used in the paper, showing the 
		number of publications, citations, 	authors, unique word 
		tokens, the average number of words in each document, and 
		the average percentage of unique words 
		repeated in a document. Note: author information is not 
		available in the last three datasets. 
	}
	\label{tbl:datasets}
	\begin{tabular}
	{
	l
	S[table-format=6.0]
	S[table-format=7.0]
	S[table-format=5.0]
	S[table-format=4.0]
	S[table-format=2.1]
	S[table-format=2.1]
	}
    \toprule
	Datasets & {Publications} & {Citations} & {Authors} & 
	{Vocabulary} & {Words/Doc} & \% {Repeat} \\
	\midrule
	ML     & 139227 & 1105462 & 43643 & 8322 & 59.4 &  23.3 \\
	M10    &  10310 &   77222 &  6423 & 2956 & 57.8 &  24.3 \\
	AvS    &  18720 &   54601 & 11898 & 4770 & 58.9 &  17.0 \\
	CS     &   3312 &    4608 & {$-$} & 3703 & 31.8 & {$-$} \\
	Cora   &   2708 &    5429 & {$-$} & 1433 & 18.2 & {$-$} \\
	PubMed &  19717 &   44335 & {$-$} & 4209 & 67.6 &  40.1 \\
	\bottomrule
	\end{tabular}
\end{table}

In addition to the manually queried datasets, we also make use of 
existing datasets from LINQS 
\citep{sen:aimag08}\footnote{\url{http://linqs.cs.umd.edu/projects/projects/lbc/}} 
to facilitate comparison with existing work. In particular, we use 
their CiteSeer, Cora and PubMed datasets. Their CiteSeer data 
consists of Computer Science publications and hence we name 
the dataset CS to remove ambiguity. Although these datasets are small, 
they are fully labelled and thus useful for clustering evaluation.
However, these three datasets do not come with additional metadata 
such as the authorship information. 
Note that the CS and Cora datasets are presented as 
Boolean matrices, \textit{i.e.}\ the word counts information is lost 
and we assume that all words in a document occur only once.
Additionally, the words have been converted to integer so they do not
convey any semantics.
Although this representation is less useful for topic modelling, we 
still use them for the sake of comparison. 
For the PubMed dataset, we recover the word counts from TF-IDF
using a simple assumption 
(see \Cref{appendix:recovering_from_tfidf}).
We present a summary of the datasets in \Cref{tbl:datasets}
and their respective categorical labels in \Cref{tbl:categories}.

\begin{table}[t!]
	\caption{Categories of the datasets.}
	\vspace{2mm}
	\centering
	\label{tbl:categories}
	\begin{tabular}{lcc}
	\toprule
	Datasets & \multicolumn{1}{c}{Classes} 
	& \multicolumn{1}{c}{Categorical Labels} 
	\\
	\midrule
	ML     &  1 & Machine Learning  \\[0.8pt]
	\hdashline
	\noalign{\vskip 1.3pt}  
	\multirow{3}{*}{M10} & \multirow{3}{*}{10} & Agriculture,  Archaeology,  Biology, 
												 Computer Science, Physics, \\ 
										      && Financial Economics,  Industrial 
												 Engineering,  Material Science, \\
											  && Petroleum Chemistry, Social Science \\[0.8pt]
	\hdashline
	\noalign{\vskip 1.3pt}  
	AvS    &  5 &  History, Religion, Physics, Chemistry, Biology  \\[0.8pt]
	\hdashline
	\noalign{\vskip 1.3pt}  
	CS     &  6 &  Agents, AI, DB, IR, ML, HCI \\[0.8pt]
	\hdashline
	\noalign{\vskip 1.3pt}  
	\multirow{2}{*}{Cora} & \multirow{2}{*}{7} & Case Based, Genetic Algorithms, 
												 Neural Networks, Theory, \\
         									  && Probabilistic Methods, 
         									     Reinforcement Learning, Rule Learning \\[0.8pt]
	\hdashline
	\noalign{\vskip 1.3pt}  
	\multirow{2}{*}{PubMed} & \multirow{2}{*}{3} & ``Diabetes Mellitus, Experimental'', \\
												&& Diabetes Mellitus Type 1, 
												   Diabetes Mellitus Type 2 \\
	\bottomrule
	\end{tabular}
\end{table}

\subsection{Data Noise Removal}
\label{subsec:preprocessing}

Here, we briefly discuss the steps taken to reduce the corrupted
entries in the CiteSeer$^\mathrm{X}$ datasets (ML, M10 and AvS).
Note that the {\it keywords} field in the publications are often 
empty and are sometimes noisy, that is, they contain irrelevant 
information such as section heading and title, which makes the 
keywords unreliable source of information as categories.
Instead, we simply treat the keywords as part of the abstracts.
We also remove the URLs from the data since they do not provide 
any additional useful information.

Moreover, the author information is not consistently presented in 
CiteSeer$^\mathrm{X}$. Some of the authors are shown with full name, 
some with first name initialised, while some others are prefixed 
with title (Prof, Dr.\ {\it etc.}). We thus standardise the author 
information by removing all title from the authors, initialising all 
first names and discarding the middle names. Although 
standardisation allows us to match up the authors, it does not solve 
the problem that different authors who have the same initial and 
last name are treated as a single author. For example, both Bruce 
Lee and Brett Lee are standardised to B Lee.  Note this 
corresponds to a whole research problem 
\citep{Han:2004:TSL:996350.996419, Han:2005:NDA:1065385.1065462} and 
hence not addressed in this paper. Occasionally, institutions are 
mistakenly treated as authors in CiteSeer$^\mathrm{X}$ data, 
example includes {\it American Mathematical Society} and 
{\it Technische Universit\"{a}t M\"{u}nchen}. In this case, we 
remove the invalid authors using a list of exclusion 
words. The list of exclusion words is presented in  
\Cref{appendix:exclusion_words}.

\subsection{Text Preprocessing}

Here, we discuss the preprocessing pipeline adopted for the 
\textit{queried} datasets (note LINQS data were already processed).
First, since publication text contains many technical terms that are 
made of multiple words, we tokenise the text using phrases (or 
collocations) instead of \textit{unigram} words. Thus, phrases like 
{\it decision tree} are treated as single token rather than two 
distinct words. 
Then, we use {\tt LingPipe} \citep{carpenter2004phrasal}%
\footnote{\url{http://alias-i.com/lingpipe/}} to extract the significant phrases 
from the respective datasets.
We refer the readers to the online tutorial%
\footnote{\url{http://alias-i.com/lingpipe/demos/tutorial/interestingPhrases/read-me.html}}
for details.
In this paper, we use the word {\it words} to mean both unigram 
words and phrases.

We then change all the words to lower case and filter out certain 
words. Words that are removed are {\it stop words}, common words and 
rare words. More specifically, we use the stop words list from
{\tt MALLET} \citep{mccallum2002mallet}.\footnote{\url{http://mallet.cs.umass.edu/}}
We define common words as words that appear in more than 18\% of the 
publications, and rare words are words that occur less than 50 times 
in each dataset.  Note that the thresholds are determined by 
inspecting the words removed. Finally, the tokenised words are 
stored as arrays of integers. We also split the datasets to 90\% 
training set for training the topic models, and 10\% test set for 
evaluations detailed in \Cref{sec:experiment}.

\section{Experiments and Results}
\label{sec:experiment}

In this section, we describe experiments that compare the SCNTM 
against several baseline topic models. The baselines are 
HDP-LDA with burstiness~\citep{Buntine:2014:ENT:2623330.2623691}, 
a nonparametric extension of the ATM, the Poisson mixed-topic link 
model (PMTLM) \citep{ZhuYGM:2013}. We also display the results for the
CNTM without the citation network for comparison purpose. 
We evaluate these models quantitatively with 
goodness-of-fit and clustering measures. 

\subsection{Experimental Settings}

In the following experiments, we initialise the concentration 
parameters $\beta$ of all PYPs to $0.1$, noting that the 
hyperparameters are updated automatically. We set the discount 
parameters $\alpha$ to $0.7$ for all PYPs corresponding to the 
``\textit{word}'' side of the SCNTM (\textit{i.e.}\ $\gamma$, $\phi$, 
$\phi'$). 
This is to induce power-law behaviour on the word 
distributions. We simply set the $\alpha$ to $0.01$ for all other 
PYPs. 
Note that the number of topics grow with data in 
nonparametric topic modelling. To prevent the learned topics from being 
too fine-grained, we set a limit to the maximum number of 
topics that can be learned. In particular, we have the number of 
topics cap at 20 for the ML dataset, 50 for the M10 dataset and 30 for the AvS 
dataset. For all the topic models, our experiments find that the 
number of topics always converges to the cap. For CS, Cora and 
PubMed datasets, we \textit{fix} the number of topics to 6, 7 and 3 
respectively for comparison against the PMTLM.

When training the topic models, we run the inference algorithm for 
2,000 iterations. For the SCNTM, the MH algorithm for the citation 
network is performed after the 1,000th iteration.  This is so the
topics can be learned from the collapsed Gibbs sampler first.  
This gives a faster learning algorithm 
and also allows us to assess the ``\textit{value-added}'' by 
the citation network to topic modelling (see \Cref{subsec:convergence}).
We repeat each experiment five times to reduce the estimation error 
of the evaluation measures.

\subsection{Estimating the Test Documents' Topic Distributions}
\label{subsec:estimate_theta}

The topic distribution $\theta'$ on the test documents is required 
to perform various evaluations on topic models. 
These topic distributions are unknown and hence need to be estimated.
Standard practice uses the first half of the text in each test document
to estimate $\theta'$, and uses the other half for evaluations.
However, since abstracts are relatively shorter compared to articles,
adopting such practice would mean there are too little text to be
used for evaluations.
Instead, we used only the words from the publication title to
estimate $\theta'$, allowing more words for evaluation.
Moreover, title is also a good indicator of topic so it is well suited
to be used in estimating $\theta'$.
The estimated $\theta'$ will be used in perplexity and clustering evaluations
below.
We note that for the clustering task, both title and abstract text 
are used in estimating $\theta'$ as there is no need to use the text
for clustering evaluation.

We briefly describe how we estimate the topic distributions $\theta'$ of 
the test documents. 
Denoting $w_{dn}$ to represent the word at position $n$ in a test document $d$, we 
\textit{independently} estimate the topic assignment $z_{dn}$ of word $w_{dn}$ by 
sampling from its predictive posterior distribution given the learned topic 
distributions $\nu$ and topic-word distributions $\phi$:
\begin{align}
p(z_{dn}=k \,|\, w_{dn}, \nu, \phi) \propto \nu_{bk} \, \phi_{kw_{dn}}~~~,
\end{align}
where $b = a_d$ if $\mathrm{significance}(a_d) = 1$, else $b = e_d$.
Note that the intermediate distributions $\phi'$ are integrated out 
(see \Cref{appendix:integrating}).

We then build the customer counts $c^{\theta_d}$ from the sampled $z$ (for 
simplicity, we set the corresponding table counts as half the customer counts). 
With these, we then estimate the document-topic distribution $\theta'$ from 
\Cref{eq:recover_vector}.

If citation network information is present, we refine the document-topic 
distribution $\theta'_d$ using the linking topic $y_{dj}$ for train document $j$ 
where $x_{dj} = 1$. The linking topic $y_{dj}$ is sampled from the estimated 
$\theta'_d$ and is added to the customer counts $c^{\theta'_d}$, which further 
updates the document-topic distribution $\theta'_d$.

Doing the above gives a sample of the document-topic distribution 
$\theta'^{(s)}_d$. We adopt a Monte Carlo approach by generating $R=500$ samples 
of $\theta'^{(s)}_d$, and calculate the Monte Carlo estimate of $\theta'_d$:
\begin{align}
\hat{\theta}'_d = \frac{\sum_s \theta'^{(s)}_d}{R} ~~.
\end{align}

\subsection{Goodness-of-fit Test}

Perplexity is a popular metric used to evaluate the goodness-of-fit 
of a topic model. Perplexity is negatively related to the likelihood 
of the observed words $\mathbf{W}$ given the model, so the lower the better:
\begin{align}
\mathrm{perplexity}(\mathbf{W}) = 
\exp\left(
-\frac{
\sum_{d=1}^D \sum_{n=1}^{N_d} \log p(w_{dn} \,|\, \theta'_d, \phi)}
{\sum_{d=1}^D N_d}
\right) ~,
\end{align}
where $p(w_{dn}|\theta'_d, \phi)$ is obtained by summing over all 
possible topics:
\begin{align}
p(w_{dn} \,|\, \theta'_d, \phi) = 
\sum_k p(w_{dn} \,|\, z_{dn}=k, \phi_k) \, p(z_{dn}=k|\theta'_d) 
= \sum_k \phi_{kw_{dn}} \theta'_{dk} ~~,
\label{eq:word_probability}
\end{align}
again noting that the distributions $\phi'$ and $\theta$ are integrated out
(see the method in \Cref{appendix:integrating}).

\begin{table}[t!]
	\centering
	\caption{
		Perplexity for the train and test documents for all datasets, lower perplexity is better.
		Note that nonparametric ATM is not performed for the last three datasets due
		to the lack of authorship information in these datasets.
		}
	\label{tbl:perplexity_combined}
	\begin{tabular}{rr@{\,\tiny$\pm$\,}lr@{\,\tiny$\pm$\,}lr@{\,\tiny$\pm$\,}lr@{\,\tiny$\pm$\,}l}
	\toprule
	  \multicolumn{1}{c}{\multirow{2}{*}{Models}}
	& \multicolumn{8}{c}{Perplexity}
	\\
	& \multicolumn{2}{c}{Train} 
	& \multicolumn{2}{c}{Test} 
	& \multicolumn{2}{c}{Train} 
	& \multicolumn{2}{c}{Test} 
	\\
	\midrule
	& \multicolumn{4}{c}{\textbf{\underline{ML}}} 
	& \multicolumn{4}{c}{\textbf{\underline{M10}}} 
	\\
	Bursty HDP-LDA 
	& $ 4904.2 $ & {\tiny $ 71.3 $} 
	& $ 4992.9 $ & {\tiny $ 65.6 $} 
	& $ 1959.4 $ & {\tiny $ 32.8 $} 
	& $ 2265.2 $ & {\tiny $ 68.2 $} 
	\\ 
	Non-parametric ATM 
	& $ 2238.2 $ & {\tiny $ 12.2 $} 
	& $ 2460.3 $ & {\tiny $ 11.3 $}
	& $ 1562.9 $ & {\tiny $ 18.1 $} 
	& $ 1814.0 $ & {\tiny $ 23.2 $}  
	\\ 
	CNTM w/o network 
	& $ 1918.2 $ & {\tiny $  4.3 $} 
	& $ 2057.6 $ & {\tiny $  3.6 $} 
	& $  912.7 $ & {\tiny $ 10.9 $} 
	& $ 1186.1 $ & {\tiny $  8.3 $} 
	\\
	SCNTM ($\eta = 0$)
	& $ \textbf{1851.8} $ & {\tiny $  8.5 $} 
	& $ \textbf{1990.8} $ & {\tiny $ 11.4 $} 
	& $ \textbf{ 824.0} $ & {\tiny $ 12.0 $} 
	& $ \textbf{1048.3} $ & {\tiny $ 21.4 $} 
	\\
	\midrule
	& \multicolumn{4}{c}{\textbf{\underline{AvS}}}
	& \multicolumn{4}{c}{\textbf{\underline{CS}}} 
	\\
	Bursty HDP-LDA 
	& $ 2460.4 $ & {\tiny $ 66.4 $} 
	& $ 2612.8 $ & {\tiny $ 91.7 $}
	& $ 1509.2 $ & {\tiny $  4.1 $} 
	& $ 1577.8 $ & {\tiny $ 33.8 $}  
	\\
	Non-parametric ATM 
	& $ 2199.7 $ & {\tiny $ 5.0 $} 
	& $ 2481.7 $ & {\tiny $ 6.1 $}   
	& \NAcell
	& \NAcell
	\\ 
	CNTM w/o network 
	& $ 1621.5 $ & {\tiny $ 19.5 $} 
	& $ 2079.4 $ & {\tiny $  2.6 $}  
	& $ 1509.4 $ & {\tiny $  4.1 $} 
	& $ 1580.2 $ & {\tiny $ 32.6 $}  
	\\
	SCNTM ($\eta = 0$)
	& $ \textbf{1620.6} $ & {\tiny $  2.2 $} 
	& $ \textbf{2028.0} $ & {\tiny $ 10.9 $}  
	& $ \textbf{1275.3} $ & {\tiny $ 14.0 $} 
	& $ \textbf{1530.8} $ & {\tiny $ 49.8 $}  
	\\
	\midrule
	& \multicolumn{4}{c}{\textbf{\underline{Cora}}}
	& \multicolumn{4}{c}{\textbf{\underline{PubMed}}} 
	\\
	Bursty HDP-LDA 
	& $ 678.1 $ & {\tiny $  2.0 $} 
	& $ 706.8 $ & {\tiny $ 17.0 $}
	& $ \textbf{299.9} $ & {\tiny $  0.2 $} 
	& $ \textbf{300.1} $ & {\tiny $  1.2 $}  
	\\
	CNTM w/o network 
	& $ 682.4 $ & {\tiny $  1.5 $} 
	& $ 702.5 $ & {\tiny $ 13.4 $}  
	& $ 301.0 $ & {\tiny $  0.2 $} 
	& $ 301.2 $ & {\tiny $  1.2 $}  
	\\
	SCNTM ($\eta = 0$)
	& $ \textbf{621.1} $ & {\tiny $  6.7 $} 
	& $ \textbf{688.0} $ & {\tiny $ 15.7 $}  
	& $ 312.3 $ & {\tiny $  1.3 $} 
	& $ 303.2 $ & {\tiny $  1.2 $} 
	\\ 	
	\bottomrule
	\end{tabular}
\end{table}

We can calculate the perplexity estimate for both the training data and test data. 
Note that the perplexity estimate is unbiased since the words used 
in estimating $\theta$ are not used for evaluation.
We present the perplexity result in \Cref{tbl:perplexity_combined}, 
showing the significantly (at 5\% significance level) 
better performance of SCNTM against the baselines on ML, M10 and AvS datasets.
For these datasets, inclusion of citation information also provides additional 
improvement for model fitting, as shown in the comparison with CNTM without 
network component. For the CS, Cora and PubMed datasets, the nonparametric ATM was not 
performed due to the lack of authorship information. We note that the results for 
other $\eta$ is not presented as they are significantly worse than $\eta=0$.
This 
is because the models are more restrictive, causing the likelihood to be worse.
We like to point out that when no author is observed, the CNTM 
is more akin to a variant of HDP-LDA which uses PYP instead of DP, 
this explains why the perplexity results are very similar.

\subsection{Document Clustering}
\label{subsubsec:clustering}

Next, we evaluate the clustering ability of the topic models.
Recall that topic models assign a topic to each word in a document, 
essentially performing a \textit{soft clustering} in which the membership is 
given by the document-topic distribution $\theta$.
For the following evaluation, we convert the soft clustering to hard 
clustering by choosing a topic that best represents the documents, 
hereafter called the \textit{dominant topic}. The dominant topic 
corresponds to the topic that has the highest proportion in a topic 
distribution. 

\begin{table}[t]
	\centering
	\caption{Comparison of clustering performance. 
			The best PMTML results are chosen for comparison, 
			from Table 2 in \cite{ZhuYGM:2013}.}
	\label{tbl:clustering}
	\begin{tabular}{rr@{\,\tiny$\pm$\,}lr@{\,\tiny$\pm$\,}lr@{\,\tiny$\pm$\,}lr@{\,\tiny$\pm$\,}l}
	\toprule
	\multicolumn{1}{c}{Models}
	& \multicolumn{2}{c}{Purity} 
	& \multicolumn{2}{c}{NMI} 
	& \multicolumn{2}{c}{Purity} 
	& \multicolumn{2}{c}{NMI} 
	\\
	\midrule
	& \multicolumn{4}{c}{\textbf{\underline{M10}}} 
	& \multicolumn{4}{c}{\textbf{\underline{AvS}}} 
	\\
	Bursty HDP-LDA 
	& $ 0.66 $ & {\tiny $ 0.02 $} 
	& $ 0.67 $ & {\tiny $ 0.01 $} 
	& $ \textbf{0.75} $ & {\tiny $ 0.03 $} 
	& $ \textbf{0.66} $ & {\tiny $ 0.01 $}
	\\ 
	Non-parametric ATM 
	& $ 0.58 $ & {\tiny $ 0.01 $} 
	& $ 0.63 $ & {\tiny $ 0.00 $} 
	& $ 0.69 $ & {\tiny $ 0.02 $} 
	& $ 0.64 $ & {\tiny $ 0.01 $}
	\\ 
	CNTM w/o network
	& $ 0.61 $ & {\tiny $ 0.04 $}  
	& $ 0.67 $ & {\tiny $ 0.01 $} 
	& $ 0.72 $ & {\tiny $ 0.03 $} 
	& $ \textbf{0.66} $ & {\tiny $ 0.01 $}
	\\ 
	SCNTM ($\eta = 0$)
	& $ 0.67 $ & {\tiny $ 0.03 $} 
	& $ 0.69 $ & {\tiny $ 0.02 $} 
	& $ 0.72 $ & {\tiny $ 0.01 $} 
	& $ \textbf{0.66} $ & {\tiny $ 0.00 $}
	\\ 
	SCNTM ($\eta = 10$)
	& $ \textbf{0.73} $ & {\tiny $ 0.02 $} 
	& $ \textbf{0.72} $ & {\tiny $ 0.01 $}  
	& $ 0.73 $ & {\tiny $ 0.01 $} 
	& $ \textbf{0.66} $ & {\tiny $ 0.01 $}
	\\
	SCNTM ($\eta = \infty$)
	& $ 0.70 $ & {\tiny $ 0.03 $} 
	& $ 0.70 $ & {\tiny $ 0.02 $}  
	& $ 0.73 $ & {\tiny $ 0.02 $} 
	& $ \textbf{0.66} $ & {\tiny $ 0.01 $}
	\\
	\midrule
	& \multicolumn{4}{c}{\textbf{\underline{CS}}} 
	& \multicolumn{4}{c}{\textbf{\underline{Cora}}} 
	\\
	PMTLM
	& \NAcell 
	& \multicolumn{1}{c}{$ 0.51$} &
	& \NAcell 
	& \multicolumn{1}{c}{$ 0.41 $}  &
	\\ 
	Bursty HDP-LDA 
	& $ 0.46 $ & {\tiny $ 0.11 $} 
	& $ 0.63 $ & {\tiny $ 0.03 $} 
	& $ 0.34 $ & {\tiny $ 0.03 $} 
	& $ 0.58 $ & {\tiny $ 0.01 $} 
	\\ 	
	CNTM w/o network
	& $ 0.51 $ & {\tiny $ 0.07 $}
	& $ 0.67 $ & {\tiny $ 0.02 $}	
	& $ 0.37 $ & {\tiny $ 0.03 $}	
	& $ 0.63 $ & {\tiny $ 0.01 $}
	\\ 
	SCNTM ($\eta = 0$)
	& $ 0.51 $ & {\tiny $ 0.08 $}	
	& $ 0.66 $ & {\tiny $ 0.02 $}
	& $ 0.39 $ & {\tiny $ 0.03 $}	
	& $ 0.63 $ & {\tiny $ 0.02 $}
	\\ 
	SCNTM ($\eta = \infty$)
	& $ \textbf{0.54} $ & {\tiny $ 0.10 $}	
	& $ \textbf{0.69} $ & {\tiny $ 0.04 $}	
	& $ \textbf{0.47} $ & {\tiny $ 0.06 $}	
	& $ \textbf{0.66} $ & {\tiny $ 0.03 $}
	\\ 
	\midrule
	& \multicolumn{4}{c}{\textbf{\underline{PubMed}}} 
	\\
	PMTLM
	& \NAcell 
	& \multicolumn{1}{c}{$ 0.27 $} & \\ 
	Bursty HDP-LDA 
	& $ \textbf{0.53} $ & {\tiny $ 0.04 $} 
	& $ \textbf{0.73} $ & {\tiny $ 0.01 $} 
	\\ 	
	CNTM w/o network
	& $ 0.47 $ & {\tiny $ 0.04 $}
	& $ 0.69 $ & {\tiny $ 0.01 $}	
	\\ 
	SCNTM ($\eta = 0$)
	& $ 0.46 $ & {\tiny $ 0.02 $}
	& $ 0.69 $ & {\tiny $ 0.01 $} 
	\\ 
	SCNTM ($\eta = \infty$)
	& $ 0.52 $ & {\tiny $ 0.01 $}
	& $ 0.72 $ & {\tiny $ 0.01 $} 	
	\\ 
	\bottomrule
	\end{tabular}
\end{table}

As mentioned in \Cref{sec:data}, for M10 and AvS datasets, 
we assume their ground truth classes correspond to the query categories used in 
creating the datasets. The ground truth classes for CS, Cora and PubMed datasets 
are provided. We evaluate the clustering performance with \textit{purity} and 
\textit{normalised mutual information} 
(NMI)
\citep{Manning:2008:IIR:1394399}. 
Purity is a simple clustering measure which can be interpreted as the proportion 
of documents correctly clustered, while NMI is an information theoretic measures 
used for clustering comparison. For ground truth classes 
$\mathcal{S} = \{s_1, \dots, s_J\}$ and obtained clusters 
$\mathcal{R} = \{r_1, \dots, r_K\}$, the purity and NMI are computed as 
\begin{align}
\mathrm{purity}(\mathcal{S}, \mathcal{R}) = 
\frac{1}{D} \sum_k \max_j | r_k \cap s_j | ~~,
&& 
\mathrm{NMI}(\mathcal{S}, \mathcal{R}) = 
\frac{2 \, I(\mathcal{S}; 
\mathcal{R})}{H(\mathcal{S})+H(\mathcal{R})} ~~,
\end{align}
where $I(\mathcal{S}; \mathcal{R})$ denotes the mutual information 
and $H(\cdot)$ denotes the entropy:
\begin{align}
I(\mathcal{S}; \mathcal{R}) = 
\sum_{k,\,j} 
\frac{| r_k \cap s_j |}{D} 
\log_2 \frac{D |r_k \cap s_j|}{| r_k | | s_j |} ~~,
&& 
H(\mathcal{R}) = - \sum_k \frac{|r_k|}{D} \log_2 \frac{|r_k|}{D} 
~~.
\end{align}

The clustering results are presented in \Cref{tbl:clustering}. 
We can see that the SCNTM greatly outperforms the PMTLM in NMI evaluation.  Note 
that for a fair comparison against PMTLM, the experiments on the CS, Cora and 
PubMed datasets are evaluated with a 10-fold cross validation. We find that 
incorporating supervision into the topic model leads to improvement on clustering 
task, as predicted. However, this is not the case for the PubMed dataset. We 
suspect this is because the publications in the PubMed dataset are highly related 
to one another so the category labels are less useful (see 
\Cref{tbl:categories}).

\section{Qualitative Analysis of Learned Topic Models}
\label{sec:qualitative_analysis}

We move on to perform qualitative analysis on the learned 
topic models in this section.
More specifically, we inspect the learned topic-word distributions, 
as well as the topics associated with the authors.
Additionally, we present a visualisation of the 
author-topic network learned by the SCNTM.

\subsection{Topical Summary of the Datasets}

By analysing the topic-word distribution $\phi_k$ for each topic $k$, 
we obtain the topical summary of the datasets. 
This is achieved by querying the top words associated with each
topic $k$ from $\phi_k$, which are learned by the SCNTM.
The top words give us an idea of what the topics are about.
In \Cref{tbl:topic_combined}, we display some major topics extracted 
and the corresponding top words. 
We note that the topic labels are manually assigned based on the top words.
For example, we find that the major topics associated with the ML dataset
are various disciplines on machine learning such as reinforcement learning
and data mining.

We did not display the topical summary for the CS, Cora and PubMed datasets. 
The reason being that the original word information is lost in the CS and Cora 
datasets since the words were converted into integers, which are not meaningful.
While for the PubMed dataset, we find that the topics are too similar to each other
and thus not interesting. This is mainly because the PubMed dataset focuses only 
on one particular topic, which is on Diabetes Mellitus.

\subsection{Analysing Authors' Research Area}

\begin{table}[t]
	\centering
	\caption{
		Topical summary for the ML, M10 and AvS datasets. The top words 
		are extracted from the topic-word distributions $\phi$ learned by SCNTM.
		}
	\label{tbl:topic_combined}
	\begin{tabular}{cc}
	\toprule
	\multicolumn{1}{c}{Topic} & \multicolumn{1}{c}{Top Words} \\
	\midrule
	& \multicolumn{1}{l}{\textbf{\,\underline{ML}}} \\
	Reinforcement Learning & reinforcement, agents, control, state, task \\
	Object Recognition & face, video, object, motion, tracking \\
	Data Mining & mining, data mining, research, patterns, knowledge \\
	SVM & kernel, support vector, training, clustering, space \\
	Speech Recognition & recognition, speech, speech recognition, audio, hidden markov\!\! \\
	\midrule
	& \multicolumn{1}{l}{\textbf{\underline{M10}}} \\
	DNA Sequencing & genes, gene, sequence, binding sites, dna \\
	Agri\-culture & soil, water, content, soils, ground \\
	Financial Market & volatility, market, models, risk, price \\
	Bayesian Modelling & bayesian, methods, models, probabilistic, estimation \\
	Quantum Theory & quantum, theory, quantum mechanics, classical, quantum field\!\! \\
	\midrule
	& \multicolumn{1}{l}{\textbf{\underline{AvS}}} \\
	Language Modelling & type, polymorphism, types, language, systems \\
	Molecular Structure & copper, protein, model, water, structure \\
	Quantum Theory & theory, quantum, model, quantum mechanics, systems \\
	Social Science & research, development, countries, information, south africa \\
	Family Well-being & children, health, research, social, women \\
	\bottomrule
	\end{tabular}
%
	\centering
	\caption{
		Major authors and their main research area. Top words are extracted
		from the topic-word distribution $\phi_k$ corresponding to the 
		dominant topic $k$ of the author.
		}
	\label{tbl:author_interest}
	\begin{tabular}{lcc}
	\toprule
	\multicolumn{1}{c}{Author}	& \multicolumn{1}{c}{Topic} & \multicolumn{1}{c}{Top Words} 
	\\
	\midrule
	D. Aerts & Quantum Theory & quantum, theory, quantum mechanics, classical 
	\\
	Y. Bengio & Neural Network & networks, learning, recurrent, neural 
	\\
	C. Boutilier & Decision Making & decision making, agents, decision, theory, agent 
	\\
	S. Thrun & Robot Learning & robot, robots, control, autonomous, learning 
	\\
	M. Baker & Financial Market & market, risk, firms, returns, financial 
	\\
	E. Segal & Gene Clustering & clustering, processes, gene expression, genes
	\\
	P. Tabuada & Control System & systems, hybrid, control systems, system, control
	\\
	L. Ingber & Statistical Mechanic & statistical, mechanics, systems, users, interactions
	\\
	\bottomrule
	\end{tabular}
\end{table}

In SCNTM, we model the author-topic distribution $\nu_i$ for each author $i$.
This allows us to analyse the topical interest of each author in a 
collection of publications. Here, we focus on the M10 dataset since it covers a 
more diverse research areas. For each author $i$, we can determine their dominant 
topic $k$ by looking for the largest topic in $\nu_i$. Knowing the dominant topic 
$k$ of the authors, we can then extract the corresponding top words from the 
topic-word distribution $\phi_k$. In \Cref{tbl:author_interest}, we display 
the dominant topic associated with several major authors and the corresponding top 
words. For instance, we can see that the author D.\,Aerts's main research area is in
quantum theory, while M.\,Baker focuses on financial markets. Again, we note that 
the topic labels are manually assigned to the authors based on the top words 
associated with their dominant topics.

\subsection{Author-topics Network Visualisation}

In addition to inspecting the topic and word distributions, we present 
a way to graphically visualise the author-topics network extracted by SCNTM, using 
\texttt{Graphviz}.\footnote{\url{http://www.graphviz.org/}}
On the ML, M10 and AvS datasets, we analyse the influential authors and 
their connections with the various topics learned by SCNTM.
The influential authors are determined based on a measure we call 
author influence, which is the sum of the $\lambda^-$ of all their publications,
\textit{i.e.}\ the influence of an author $i$ is
$\sum_d \lambda^-_d \, I(a_d = i)$.
Note that $a_d$ denotes the author of document $d$,
and $I(\cdot)$ is the indicator function, as previously defined.

\begin{figure}[t]
	\begin{center}
	\setlength{\fboxsep}{0pt}%
	\setlength{\fboxrule}{2pt}%
	\fbox{%
		\includegraphics[width=0.98\columnwidth]{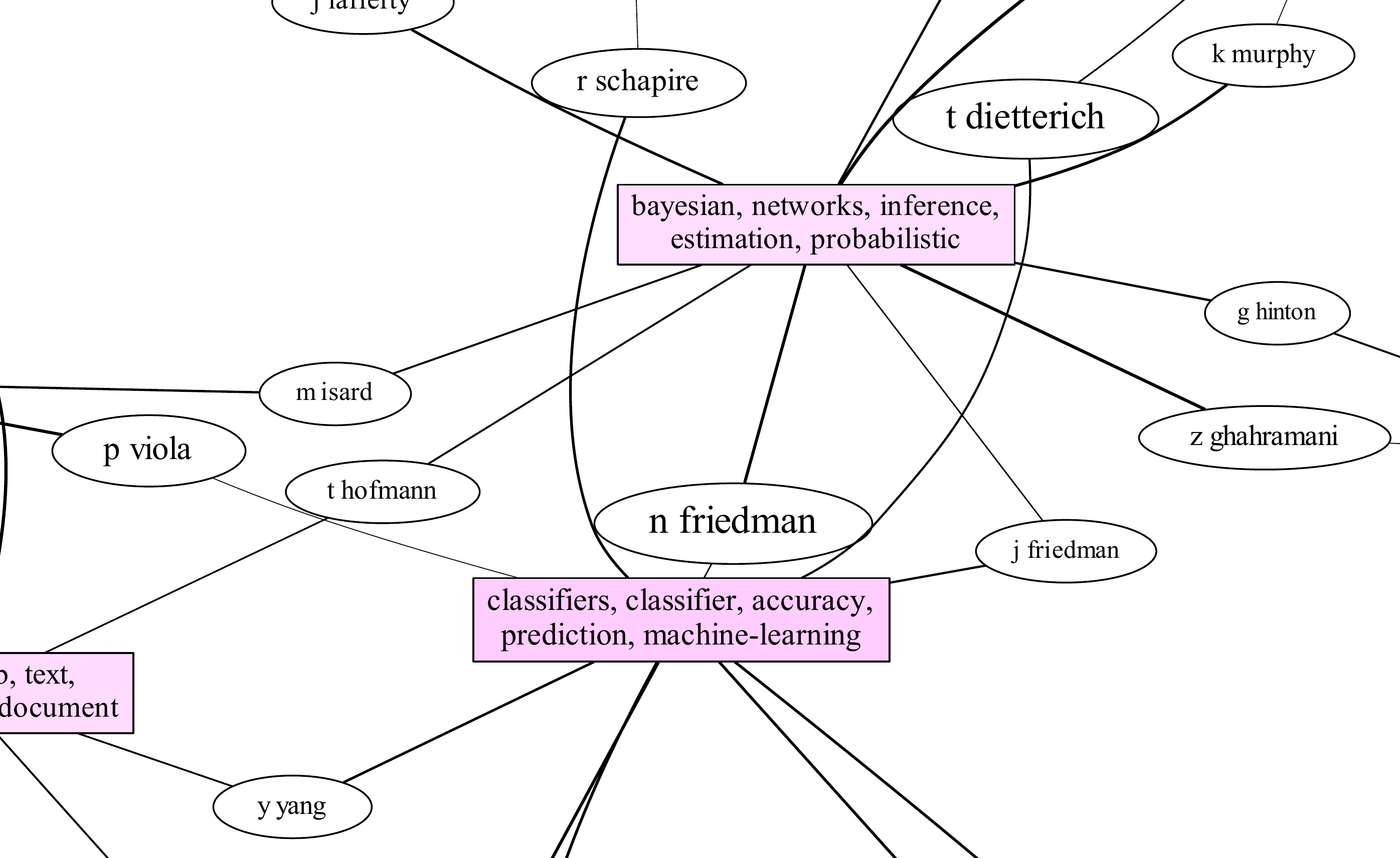}%
	}
	\caption{
		Snapshot of the author-topics network from the ML dataset. 
		The pink rectangles represent the learned topics, 
		their intensity (pinkness) corresponds to the topic proportion. 
		The ellipses represent the authors, 
		their size corresponds to the author's influence in the corpus. 
		The strength of the connections are given by the lines' thickness.
	}
	\label{fig:ML_graph_cut}
	\vspace{-4mm}
	\end{center}
\end{figure}

\Cref{fig:ML_graph_cut} shows a snapshot of the 
author-topics network of the ML dataset.
The pink rectangles in the snapshot represent the topics learned by SCNTM, 
showing the top words of the associated topics.
The colour intensity (pinkness) of the rectangle shows the relative 
weight of the topics in the corpus.
Connected to the rectangles are ellipses representing the authors, 
their size is determined by their corresponding author influence in the corpus.
For each author, the thickness of the line connecting to a topic shows the 
relative weight of the topic.
Note that not all connections are shown, some of the weak connections are dropped
to create a neater diagram.
In \Cref{fig:ML_graph_cut}, we can see that Z.\,Ghahramani 
works mainly in the area of Bayesian inference, 
as illustrated by the strong connection to 
the topic with top words ``bayesian, networks, inference, estimation, probabilistic".
While N.\,Friedman works in both Bayesian inference and machine learning classification, 
though with a greater proportion in Bayesian inference.
Due to the large size of the plots, we present 
online\footnote{\url{https://drive.google.com/folderview?id=0B74l2KFRFZJmVXdmbkc3UlpUbzA} \\
(please download and view with a web browser for best quality)} 
the full visualisation of the author-topics network learned from the CiteSeer$^\mathrm{X}$ datasets.

\section{Diagnostics}
\label{sec:diagnostic}

In this section, we perform some diagnostic tests for the SCNTM. We assess the 
convergence of the MCMC algorithm associated with SCNTM and inspect the counts 
associated with the PYP for the document-topic distributions. Finally, we also 
present a discussion on the running time of the SCNTM.

\subsection{Convergence Analysis}
\label{subsec:convergence}

\begin{figure}[t]
\begin{center}
  \centering
  \includegraphics[width=0.7\columnwidth]{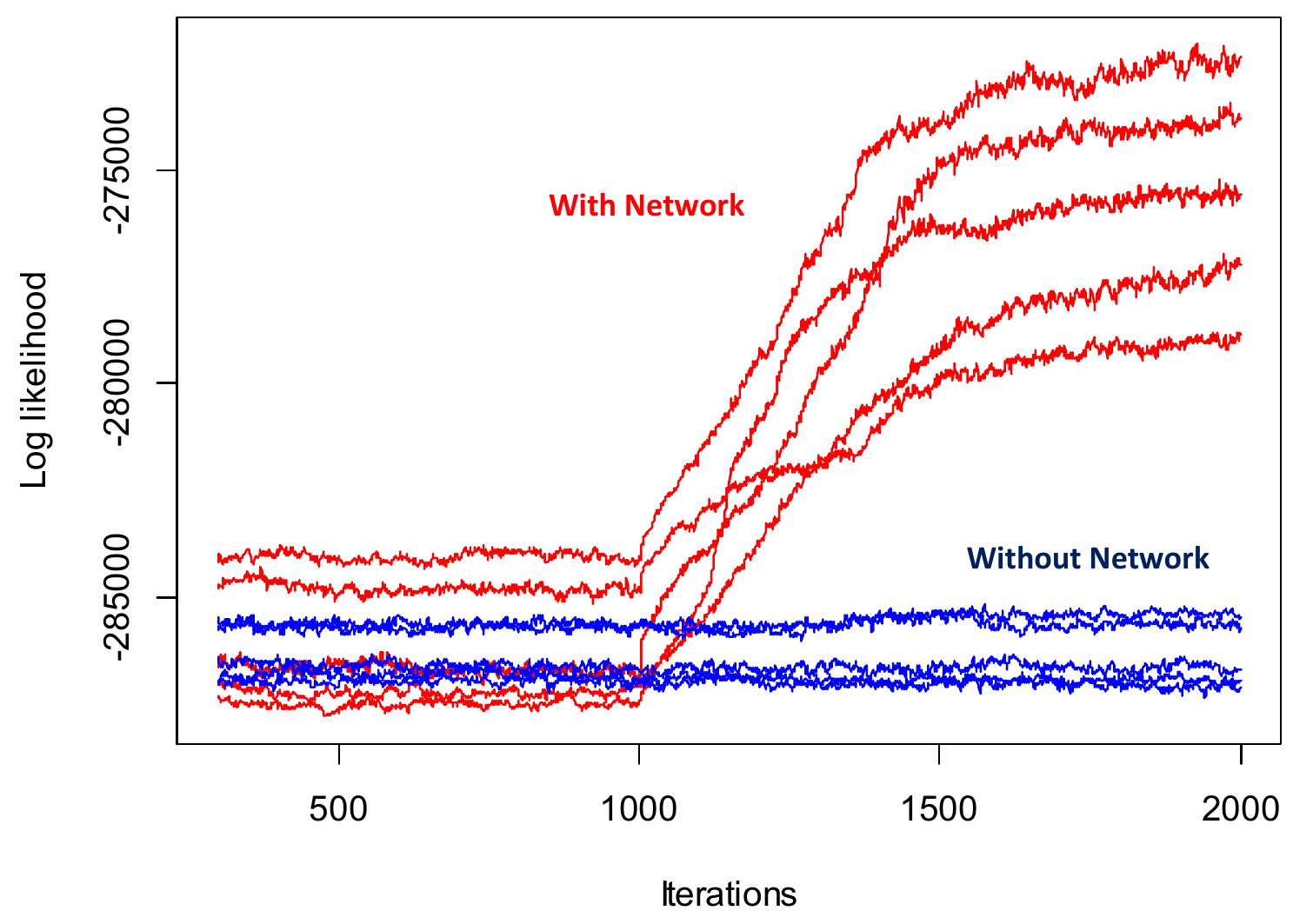}
  \caption{(Coloured) Training word log likelihood \textit{vs} iterations during 
      training of the CNTM with and without the network component. 
      The red lines show the log likelihoods of the SCNTM with the 
      citation network while the blue lines represent the SCNTM 
      without the citation network. The five runs are from five
      different folds of the Cora dataset. 
      }
  \label{fig:likelihood}
  \vspace{-4mm}
\end{center}
\end{figure}

It is important to assess the convergence of an MCMC algorithm to make sure 
that the algorithm is not prematurely terminated. In \Cref{fig:likelihood}, 
we show the time series plot of the training word log likelihood 
$\sum_{d,n} \log(p(w_{dn} \,|\,  z_{dn}, \phi'))$ corresponds to the SCNTM trained with 
and without the network information. Recall that for SCNTM, the sampler for the 
topic model is first performed for 1,000 iterations before running the full MCMC 
algorithm. From \Cref{fig:likelihood}, we can clearly see that the sampler 
converges quickly. For SCNTM, it is interesting to see that the log likelihood 
improves significantly once the network information is used for training (red 
lines), suggesting that the citation information is useful. Additionally, we like 
to note that the acceptance rate of the MH algorithm for the citation network 
averages about 95\%, which is very high, suggesting that the proposed MH 
algorithm is~effective.

\subsection{Inspecting Document-topic Hierarchy}
\label{subsec:modelling_topic_hierarchy}

As previously mentioned, modelling the document-topic hierarchy allows us to
balance the contribution of text information and citation information
toward topic modelling. 
In this section, we inspect the customer and table counts associated 
with the document-topic distributions $\theta'$ and $\theta$ to give
an insight on how the above modelling works.
We first note that the number of words in a document tend to be higher
than the number of citations.

We illustrate with an example from the ML dataset.  We look at
the 600th document, which contains $84$ words but only $4$ citations.
The words are assigned to two topics and we have $c_1^\theta = 53$ and
$c_2^\theta = 31$. These customer counts are contributed to $\theta'$
by way of the corresponding table counts $t_1^\theta = 37$ and 
$t_2^\theta = 20$. 
The citations contribute counts directly to $\theta'$, in this case, 
three of the citations are assigned the first topic while another one 
is assigned to the second topic.
The customer count for $\theta'$ is the sum of the table counts from $\theta$
and the counts from citations.
Thus, $c_1^{\theta'} = 37 + 3 = 40$ and $c_2^{\theta'} = 20 + 1 = 21$.
Note that the counts from $\theta'$ are used to determine the topic composition 
of the document. 
By modelling the document-topic hierarchy, 
we have effectively diluted the influence of text information.  This is 
essential to counter the higher number of words compared to citations.

\subsection{Computation Complexity}
\label{subsec:computation_complexity}

Finally, we briefly discuss the computational complexity of 
the proposed MCMC algorithm for the SCNTM. Although we did 
not particularly optimise our implementation for algorithm 
speed, the algorithm is of linear time with the number of 
words, the number of citations and the number of topics. 
All implementations are written in \texttt{Java}.

We implemented a general sampling framework that works with arbitrary PYP network,
this allows us to test various PYP topic models with ease and without spending too 
much time in coding. However, having a general framework for PYP topic models 
means it is harder to optimise the implementation, thus it performs slower than 
existing implementations (such as 
\texttt{hca}\footnote{\url{http://mloss.org/software/view/527/}}). 
Nevertheless, the running time is linear with the number of words in the corpus 
and the number of topics, and constant time with the number of citations.

A na\"{\i}ve implementation of the MH algorithm for the citation network would be of 
polynomial time, due to the calculation of the double summation in the posterior. 
However, with caching and reformulation of the double summation, we can evaluate 
the posterior in linear time. Our implementation of the MH algorithm is linear (in 
time) with the number of citations and the number of topics,  and it is constant 
time with respect to the number of words. The MCMC algorithm is constant time with 
respect to the number of authors.

\Cref{tbl:running_time} shows the average time taken to perform the MCMC 
algorithm for 2000 iterations. All the experiments were performed with a machine 
having \texttt{Intel(R) Core(TM) i7 CPU @ 3.20GHz} (though only 1 processor was 
used) and 24 Gb RAM.

\begin{table}[t!]
	\centering
	\caption{
		Time taken to perform 2,000 iterations of the MCMC algorithm given the statistics of the datasets. 
		The reported SCNTM run time corresponds to $\eta = \infty$.
	}
	\label{tbl:running_time}
	\begin{tabular}{
	c
	S[table-format=7.0]
	S[table-format=7.0]
	S[table-format=2.0]
	S[table-format=5.0]
	}
	\toprule
	{Datasets}
	& {Total Words}
	& {Citations}
	& {Number of Topics}
	& {Time [mins]}
	\\
	\midrule
	ML		&	8270084	&	1105462	&	20 & 16444 
	\\ 
	M10		&	 595918	&	  77222	&	50 &  1845 
	\\ 
	AvS		&	1102608	&	  54601	&	30 &  2092 
	\\ 
	CS		&	 105322	&	   4608	&	 6 &    43 
	\\ 
	Cora	&	  49286	&	   5429	&	 7 &    26 
	\\
	PubMed	&	1332869	&	  44335	&	 3 &   397 
	\\
	\bottomrule
	\end{tabular}
\end{table}

\section{Conclusions}
\label{sec:conclusion}

In this paper, we have proposed the Supervised Citation Network Topic Model 
(SCNTM) as an extension of our previous work \citep{LimBuntineCNTM} to jointly 
model research publications and their citation network. The SCNTM makes use of the 
author information as well as the categorical labels associated with each document 
for supervised learning. The SCNTM performs text modelling with a hierarchical PYP 
topic model and models the citations with the Poisson distribution given the 
learned topic distributions. We also proposed a novel learning algorithm for the 
SCNTM, which exploits the conjugacy of the Dirichlet distribution and the 
Multinomial distribution, allowing the sampling of the citation networks to be of 
similar form to the collapsed sampler of a topic model. As discussed, our learning 
algorithm is intuitive and easy to implement.

The SCNTM offers substantial performance improvement over previous work 
\citep{ZhuYGM:2013}. On three CiteSeer$^\mathrm{X}$ datasets and three existing 
and publicly available datasets, we demonstrate the improvement of joint topic and 
network modelling in terms of model fitting and clustering evaluation. 
Additionally, incorporating supervision into the SCNTM provides further 
improvement on the clustering task. Analysing the learned topic models let us 
extract useful information on the corpora, for instance, we can inspect the 
learned topics associated with the documents and examine the research interest of 
the authors. We also visualise the author-topic network learned by the SCNTM, 
which allows us to have a quick look at the connection between the authors by way 
of their research areas.

\clearpage

\acks{NICTA is funded by the Australian Government through the 
Department of Communications and the Australian Research Council 
through the ICT Centre of Excellence Program. The authors wish to 
thank CiteSeer$^\mathrm{X}$ for providing the data.}


\appendix

\section{Delta Method Approximation}
\label{appendix:delta_method}

We employ the Delta Method to show that
\begin{align}
\int q(\theta)\,\exp(-g(\theta))\,\mathrm{d}\theta
\approx \exp(-g(\hat\theta)) \int q(\theta)\,\mathrm{d}\theta
&&
\mathrm{for\ small\ } g(\hat\theta) ~~,
\end{align}
where $\hat\theta$ is the expected value according to a distribution
proportional to $q(\theta)$, more specifically, define $p(\theta)$ as the 
probability density of $\theta$, we have
\begin{align}
\hat{\theta} = \mathbb{E}[\theta] = \int \theta \, p(\theta) \, \mathrm{d}\theta ~~,
&&
q(\theta) = \mathrm{constant} \times p(\theta) ~~.
\end{align}

\noindent
First we note that the Taylor expansion for a function $h(\theta) = \exp(-g(\theta))$ at 
$\hat\theta$ is
\begin{align}
h(\theta) 
= \sum_{n=0}^\infty \frac{1}{n!} \left( h^{(n)}(\hat{\theta}) \right) (\theta - \hat{\theta})^n
~~,
\label{eq:taylor_expansion}
\end{align}
where $h^{(n)}(\hat{\theta})$ denotes the $n$-th derivative of $h(\cdot)$ evaluated at 
$\hat{\theta}$:
\begin{align}
h^{(n)}(\hat{\theta}) 
= \left( - g'(\hat{\theta}) \right)^n \, h(\hat{\theta}) ~~.
\end{align}

\noindent
Multiply \Cref{eq:taylor_expansion} with $q(\theta)$ and integrating gives 
\begin{align}
\int q(\theta) \, h(\theta) \, \mathrm{d}\theta
& = \sum_{n=0}^\infty \frac{1}{n!} \left( h^{(n)}(\hat{\theta}) \right) 
\int q(\theta) \, (\theta - \hat{\theta})^n \, \mathrm{d}\theta
\nonumber
\\
& = \sum_{n=0}^\infty \frac{1}{n!} \left( - g'(\hat{\theta}) \right)^n 
\int q(\theta) \, (\theta - \hat{\theta})^n \, \mathrm{d}\theta
~~.
\label{eq:delta_method1}
\end{align}

\noindent
Since $g(\hat{\theta})$ is small, the term $\left( - g'(\hat{\theta}) \right)^n$ becomes 
exponentially smaller as $n$ increases.
Here we let $\left( - g'(\hat{\theta}) \right)^n \approx 0$ for $n \geq 2$.
Hence, continuing from \Cref{eq:delta_method1}:
\begin{align}
\int q(\theta) \, h(\theta) \, \mathrm{d}\theta
& \, \approx \, h(\hat\theta) \int q(\theta) \, \mathrm{d}\theta
+ \left( - g'(\hat{\theta}) \right) h(\hat{\theta}) \, 
\underbrace{\int q(\theta) \, (\theta - \hat{\theta}) \, \mathrm{d}\theta}_0
\nonumber
\\
& \, \approx \, h(\hat\theta) \int q(\theta) \, \mathrm{d}\theta  ~~~~~~.
\end{align}

\section[Keywords for Querying the CiteSeerX Datasets]{Keywords for Querying the CiteSeer$^\mathrm{X}$ Datasets}
\label{appendix:keywords}
\vspace{2mm}

1. For ML dataset:

\textbf{Machine Learning:}
\textit{machine learning, neural network, pattern recognition, indexing term, 
support vector machine, 
learning algorithm, computer vision, face recognition,
feature extraction, image processing, high dimensionality, image segmentation,
pattern classification, real time, feature space, decision tree,
principal component analysis, feature selection, backpropagation, edge detection,
object recognition, maximum likelihood, statistical learning theory, supervised learning,
reinforcement learning, radial basis function, support vector, em algorithm,
self organization, image analysis, hidden markov model, artificial neural network,
independent component analysis, genetic algorithm, statistical model, 
dimensional reduction, indexation, unsupervised learning, gradient descent,
large scale, maximum likelihood estimate, statistical pattern recognition,
cluster algorithm, markov random field, error rate, optimization problem,
satisfiability, high dimensional data, mobile robot, nearest neighbour,
image sequence, neural net, speech recognition, classification accuracy,
diginal image processing, factor analysis, wavelet transform, local minima,
probability distribution, back propagation, parameter estimation, probabilistic model,
feature vector, face detection, objective function, signal processing,
degree of freedom, scene analysis, efficient algorithm, computer simulation,
facial expression, learning problem, machine vision, dynamic system, 
bayesian network, mutual information, missing value, image database,
character recognition, dynamic program, finite mixture model, linear discriminate analysis,
image retrieval, incomplete data, kernel method, image representation,
computational complexity, texture feature, learning method, prior knowledge,
expectation maximization, cost function, multi layer perceptron,
iterated reweighted least square, data mining}.\\

\noindent
2. For M10 dataset:

 \textbf{Biology:}
\textit{enzyme, gene expression, amino acid, escherichia coli, transcription factor,
nucleotides, dna sequence, saccharomyces cerevisiae, plasma membrane, embryonics}.

 \textbf{Computer Science:}
\textit{neural network, genetic algorithm, machine learning, information retrieval, data mining, 
computer vision, artificial intelligent, optimization problem, support vector machine, feature selection}.

 \textbf{Social Science:}
\textit{ developing country, higher education, decision making, health care, high school, 
social capital, social science, public health, public policy, social support}.

 \textbf{Financial Economics:}
\textit{stock returns, interest rate, stock market, stock price, exchange rate,
asset prices, capital market, financial market, option pricing, cash flow}.

 \textbf{Material Science:}
\textit{microstructures, mechanical property, grain boundary, 
transmission electron microscopy, composite material,
materials science, titanium, silica, differential scanning calorimetry, tensile properties}.

 \textbf{Physics:}
\textit{magnetic field, quantum mechanics, field theory, black hole, kinetics,
string theory, elementary particles, quantum field theory, space time, star formation}.

 \textbf{Petroleum Chemistry:}
\textit{fly ash, diesel fuel, methane, methyl ester, diesel engine,
natural gas, pulverized coal, crude oil, fluidized bed, activated carbon}.

 \textbf{Industrial Engineering:}
\textit{power system, construction industry, induction motor, power converter, control system,
voltage source inverter, permanent magnet, digital signal processor, sensorless control, field oriented control}.

 \textbf{Archaeology:}
\textit{radiocarbon dating, iron age, bronze age, late pleistocene, middle stone age,
upper paleolithic, ancient dna, early holocene, human evolution, late holocene}.

 \textbf{Agriculture:}
\textit{irrigation water, soil water, water stress, drip irrigation, grain yield,
crop yield, growing season, soil profile, soil salinity, crop production}\\

\noindent
3. For AvS dataset:	

 \textbf{History:}
\textit{nineteeth century, cold war, south africa, foreign policy, civil war,
world war ii, latin america, western europe, vietnam, middle east}.

 \textbf{Religion:}
\textit{social support, foster care, child welfare, human nature, early intervention, 
gender difference, sexual abuse, young adult, self esteem, social services}.

 \textbf{Physics:}
\textit{magnetic field, quantum mechanics, string theory, field theory, numerical simulation, 
black hole, thermodynamics, phase transition, electric field, gauge theory}.

 \textbf{Chemistry:}
\textit{crystal structure, mass spectrometry, copper, aqueous solution, binding site, 
hydrogen bond, oxidant stress, free radical, liquid chromatography, organic compound}.

 \textbf{Biology:}
\textit{genetics, enzyme, gene expression, polymorphism, nucleotides, 
dna sequence, saccharomyces cerevisiae, cell cycle, plasma membrane, embryonics}.

%
%
%
%
%
%
%
%
%

\section{Recovering Word Counts from TF-IDF}
\label{appendix:recovering_from_tfidf}

The PubMed dataset \citep{sen:aimag08} was preprocessed to TF-IDF (term frequency-inverse document 
frequency) format, \textit{i.e.}\ the raw word count information is lost.
Here, we describe how we recover the word count information, using a simple and reasonable assumption -- that the least occurring words in a document only occur once.

We denote $t_{dw}$ as the TF-IDF for word $w$ in document $d$, 
$f_{dw}$ as the corresponding term frequency (TF), 
and $i_w$ as the inverse document frequency (IDF) for word $w$.
Our aim is to recover the word counts $c_{dw}$ given the TF-IDF.
TF-IDF is computed\footnote{Note that there are multiple ways to define a TF-IDF in practice. 
The specific TF-IDF formula used by the PubMed dataset was determined \textit{via} trial-and-error and elimination.} 
as
\begin{align}
t_{dw} = f_{dw} \times i_w ~~,
&&
f_{dw} = \frac{c_{dw}}{\sum_w c_{dw}} ~~,
&&
i_w =  \log \frac{\sum_d 1}{\sum_d I(c_{dw} > 0)} ~~,
\label{eq:tf-idf}
\end{align}
where $I(\cdot)$ is the indicator function.

We note that $I(c_{dw} > 0) = I(t_{dw} > 0)$ since the TF-IDF for a word $w$ 
is positive if and only if the corresponding word count is positive.
This allows us to compute the IDF $i_w$ easily from \Cref{eq:tf-idf}.
We can then determine the TF:
\begin{align}
f_{dw} 
& = t_{dw} / i_w
\nonumber
\\
& = t_{dw} \times \left( \log \frac{\sum_d 1}{\sum_d I(t_{dw} > 0)}  \right)^{-1} ~~.
\end{align}

Now we are left with computing $c_{dw}$ given the $f_{dw}$, however, we can obtain infinitely many 
solutions since we can always multiply $c_{dw}$ by a constant and get the same $f_{dw}$.
Fortunately, since we are working with natural language, it is reasonable to assume that the least 
occurring words in a document only occur once, or mathematically,
\begin{align}
c_{dw} = 1 ~~~~~~~~~~ \mathrm{for\ \ } w = \argmin_w f_{dw} ~~.
\end{align}

Thus we can work out the normaliser $\sum_w c_{dw}$ and recover the word counts for all words in all 
documents.
\begin{align}
\textstyle
\sum_w c_{dw} 
\displaystyle
= \frac{1}{\min_w f_{dw}} ~~,
&&
\textstyle
c_{dw} = f_{dw} \times \sum_w c_{dw} ~~.
\end{align}

\section{Exclusion Words to Detect Invalid Authors}
\label{appendix:exclusion_words}

Below is a list of words we use to filter out invalid authors during preprocessing step:

\noindent
\textit{society, university, universit\"{a}t, universitat, author, advisor, 
acknowledgement, video, mathematik, abstract, industrial, review, example, department, information,
enterprises, informatik, laboratory, introduction, encyclopedia, algorithm, section, available}

\section{Integrating Out Probability Distributions}
\label{appendix:integrating}


Here, we show how to integrate out probability distributions using the expectation of
a~PYP:
\begin{align}
p(w_{dn}|z_{dn}=k, \phi_k) & = \int_{\phi'_{dk}} p(w_{dn}, \phi'_{dk}|z_{dn}, \phi_k) 
\nonumber
\\
& = \int_{\phi'_{dk}} p(w_{dn}|z_{dn}, \phi'_{dk}) \, p(\phi'_{dk}|\phi_k) 
\nonumber
\\
& = \int_{\phi'_{dk}} \phi'_{dkw_{dn}} \, p(\phi'_{dk}|\phi_k) 
\nonumber
\\
& = \mathbb{E}[\phi'_{dkw_{dn}}|\phi_k] 
\nonumber
\\
& = \phi_{kw_{dn}} ~~~~~~~~~~~~,
\label{eq:expectation}
\end{align}
where $\mathbb{E}[\cdot]$ denotes the expectation value. We note that the last step 
(\Cref{eq:expectation})
follows from the fact that the expected value of a PYP is the probability vector corresponding to the
base distribution of the PYP (when the base distribution is a 
probability distribution).
A similar approach can be taken to integrate out the $\theta$ in \Cref{eq:word_probability}.


\bibliography{biblio}

\end{document}